\documentclass[reprint,twocolumn,aps,prb,showpacs]{revtex4-1}

\usepackage{amsmath}
\usepackage{amssymb}
\usepackage{graphicx}
\usepackage{dcolumn}
\usepackage{bm}
\usepackage[colorlinks=true, allcolors=blue]{hyperref}
\usepackage{epstopdf}
\usepackage[utf8]{inputenc}
\usepackage{amssymb}

\begin{document}

\title{Electromagnetic response of cuprate superconductors with coexisting electronic nematicity}

\author{Zhangkai Cao$^{\rm a}$, Xingyu Ma$^{\rm a}$, Huaiming Guo$^{\rm b}$, and Shiping Feng$^{\rm a}$}
\thanks{Corresponding author. E-mail address: spfeng@bnu.edu.cn}

\affiliation{$^{\rm a}$Department of Physics, Beijing Normal University, Beijing 100875, China}

\affiliation{$^{\rm b}$School of Physics, Beihang University, Beijing 100191, China}

\begin{abstract}
The electronically nematic order has emerged as a key feature of cuprate superconductors,
however, its correlation with the fundamental properties such as the electromagnetic response
remains unclear. Here the nematic-order state strength dependence of the electromagnetic
response in cuprate superconductors is investigated within the framework of the
kinetic-energy-driven superconductivity. It is shown that a significant anisotropy of the
electromagnetic response is caused by the electronic nematicity. In particular, in addition
to the pure d-wave component of the superconducting gap, the pure s-wave component of the
superconducting gap is generated by the electronic nematicity, therefore there is a coexistence
and competition of the d-wave component and the s-wave component. This coexistence and
competition leads to that the maximal condensation energy appears at around the optimal
strength of the electronic nematicity, and then decreases in both the weak and strong strength
regions, which in turn induces the enhancement of superconductivity, and gives rise to the
dome-like shape of the nematic-order state strength dependence of the superfluid density.
\end{abstract}


\maketitle


\section{Introduction}\label{Introduction}

The parent compound of cuprate superconductors is identified as a Mott insulator
\cite{Bednorz86,Anderson87,Kastner98}, in which the lack of conduction arises from anomalously
strong electron correlation. Superconductivity then is obtained by adding charge carriers to
this insulating parent compound \cite{Bednorz86} with the superconducting (SC) transition
temperature $T_{\rm c}$ that takes a dome-like shape with the underdoped and overdoped regimes
on each side of the optimal doping, where $T_{\rm c}$ reaches its maximum \cite{Drozdov18}. This
marked transformation of the electronic states thus reflects a basic fact that the same electron
correlation that leads to the Mott insulating state also generates superconductivity
\cite{Anderson87}. The key structural element of all cuprate superconductors is the set of the
square-lattice copper-oxide (ab) plane \cite{Kastner98}, and then the strongly correlated motion
of the electrons confined to the copper-oxide planes has been confirmed experimentally by the
incoherent charge-transport along the inter-plane direction \cite{Cooper94,Takenaka94}. This is
why the SC mechanism of superconductivity and processes responsible for the exotic features can
be found in the physics of this plane. After intensive investigations over more than three
decades, it has become clear that cuprate superconductors are among the most complicated systems
studied in condensed matter physics. The complications arise mainly from that apart from the
emergence of superconductivity, the strongly correlated motion of the electrons in the
copper-oxide plane also induces a variety of spontaneous symmetry-breaking orders
\cite{Vishik18,Comin16,Fradkin15,Kivelson19,Vojta09,Fradkin10,Fernandes19}, and then a
characteristic feature in the complicated phase diagram of cuprate superconductors is the
coexistence and intertwinement of these spontaneous symmetry-breaking orders with
superconductivity. In this case, it is widely believed that the understanding of the nature of
the coexistence and intertwinement of spontaneous symmetry-breaking orders with superconductivity
in cuprate superconductors is thought to be key to understanding the high-$T_{\rm c}$ phenomenon
in general.

Among these spontaneous symmetry-breaking orders, the ordered state which most evidently displays
the signature for the rotation-symmetry breaking of the square lattice underlying the copper-oxide
plane is the nematic-order state \cite{Fradkin15,Kivelson19,Vojta09,Fradkin10,Fernandes19}. By
virtue of systematic studies using various measurement techniques, the detailed information on
the nematic-order state has been available now
\cite{Fradkin15,Kivelson19,Vojta09,Fradkin10,Fernandes19}, where an agreement has emerged that
the electronic nematicity induces the anisotropic features in both the normal- and SC-states. In
particular, the experimental data detected from angle-resolved photoemission spectroscopy (ARPES)
\cite{Nakata18}, scanning tunneling microscopy (STM) \cite{Lawler10,Fujita14,Zheng17,Fujita19},
and electronic Raman scattering \cite{Loret19} indicate that the electronic structure is
inequivalent between the $k_{x}$ and $k_{y}$ directions in momentum space. The magnetic torque
measurements \cite{Hinkov08,Sato17,Daou10,Taillefer15,Wang21} show the anisotropic spin
excitation spectrum, while the resistivity anisotropy has been observed in transport experiments
\cite{Ando02,Wu17}. On the other hand, the elastoresistance measurements of the susceptibility
shows an anomaly at around the pseudogap crossover temperature \cite{Ishida20}, evidencing the
existence of a nematic phase transition and its quantum critical point. However, this nematic
quantum critical point is shifted from the optimal composition, indicating a link to
superconductivity as well as the exotic transport behavior in the strange-metal phase of cuprtae
superconductors \cite{Ando02,Wu17}. Moreover, the evolution of the characteristic energy of the
nematic-order state with doping has been studied experimentally in the entire SC phase
\cite{Fujita19}, where measured on the samples whose doping spans the pseudogap regime, this
characteristic energy and pseudogap energy are, within the experimental error, identical. These
experimental observations thus show that the electronically nematic order emerged as a key
features of cuprate superconductors has high impacts on various properties, while such an aspect
should be also reflected in the electromagnetic response.

The electromagnetic response yields essential information, both on the condensate as well as on
the quasiparticle excitations \cite{Schrieffer64,Bonn96,Sonier00,Basov05,Sonier16}. This follows
a basic fact that superconductivity is characterized by the exactly zero electrical resistance
and expulsion of magnetic fields occurring in superconductors when cooled below $T_{\rm c}$. The
later remarkable phenomenon is so-called Meissner effect \cite{Schrieffer64}, i.e., a
superconductor is placed in an external magnetic field ${\rm B}$, when this external magnetic
field ${\rm B}$ is smaller than the upper critical field ${\rm B}_{\rm c}$, the external
magnetic field ${\rm B}$ penetrates only to a penetration-depth $\lambda$ (few hundred nm for
cuprate superconductors at zero temperature) and is excluded from the main body of the system.
This magnetic-field penetration-depth is a fundamental parameter of superconductors, and
provides a direct measurement of the superfluid density
$\rho_{\rm s}$ ($\rho_{\rm s}\equiv\lambda^{-2}$)
\cite{Schrieffer64,Bonn96,Sonier00,Basov05,Sonier16}. The superfluid density is proportional to
the squared amplitude of the macroscopic wave function, and therefore describes the SC
quasiparticles \cite{Schrieffer64}. On the other hand, the layered crystal structure gives rise
to a strong structure anisotropy of cuprate superconductors, and then both the in-plane and
inter-plane electromagnetic responses have been observed experimentally. The former one is
characterized by the ab-plane magnetic-field penetration-depth (then the ab-plane superfluid
density), whereas the latter one is related to the magnetic-field-penetration (then the
superfluid density) in the c-axis direction \cite{Hosseini98,Hosseini04}. In this paper we
concentrate on the in-plane electromagnetic response only and do not consider c-axis properties,
which can be discussed, e.g., by taking into account hopping between adjacent copper-oxide
planes within the tunneling Hamiltonian approach. In the early experimental observations, the
main features of the {\it in-plane} electromagnetic response in cuprate superconductors have
been identified for all the temperature $T\leq T_{c}$ throughout the SC dome, and can be
summarized as: (i) the magnetic-field screening is observed to follow an exponential field decay
\cite{Jackson00,Khasanov04,Suter04}, in support of a local (London-type) nature of the
electrodynamic response \cite{Schrieffer64}; (ii) the magnetic-field penetration-depth is found
to be a generally linear temperature dependence at low temperatures, however, close to the
extremely low temperatures, this dependence becomes nonlinear
\cite{Bozovic16,Brewer15,Deepwell13,Broun07,Kim03,Panagopoulos99,Lee96,Hardy93}; (iii) the
superfluid density $\rho_{\rm s}$ exhibits a dome-like shape of the doping dependence
\cite{Lemberger11,Liang06,Bernhard01}, which in turn gives rise to the dome-like shape of the
doping dependence of $T_{\rm c}$. Later, the particularly large in-plane anisotropy of the
magnetic-field penetration-depth (then the superfluid density) in YBa$_{2}$Cu$_{3}$O$_{6+x}$
has been observed experimentally \cite{Zhang94,Basov95,Sun95,Barnea04,Kiefl10}.
YBa$_{2}$Cu$_{3}$O$_{6+x}$ has a orthorhombic crystal structure associated with the presence
of the copper-oxide chain. This copper-oxide chain is a unique feature of
YBa$_{2}$Cu$_{3}$O$_{6+x}$ which distinguishes it from other cuprate superconductors. In this
case, it has been argued that any small and weak temperature dependence of the anisotropy seen
in high temperature can be identical as the direct consequence of the orthorhombic crystal
structure, while any large magnitude, strongly temperature dependent enhancement of the
anisotropy that occurs below a well-defined crossover (transition) temperature can be plausibly
connected with the onset of the electronically nematic order \cite{Fradkin10}. Although the
effect from the orthorhombic crystal structure on the anisotropy of the electromagnetic response
in YBa$_{2}$Cu$_{3}$O$_{6+x}$ is still not fully understood on the microscopic level, it is
possible that the intrinsic aspects of the electromagnetic response in YBa$_{2}$Cu$_{3}$O$_{6+x}$
with coexisting electronic nematicity are masked by the orthorhombic crystal structure. On the
other hand, although the experimental data of the anisotropy of the superfluid density response
for La$_{2-x}$Sr$_{x}$CuO$_{4}$ are still lacking to date, the anisotropy of the electronic
structure in La$_{2-x}$Sr$_{x}$CuO$_{4}$ has been observed experimentally
\cite{Zhang18,Razzoli17}. In particular, the in-plane anisotropy of the superfluid density response
has been detected experimentally in Bi$_{2}$Sr$_{2}$CaCu$_{2}$O$_{8+\delta}$ with the mild
enhancement of the magnitude of the $\hat{a}$-axis magnetic-field penetration-depth
\cite{Quijada99}. As in the case of the crystal structure for La$_{2-x}$Sr$_{x}$CuO$_{4}$, this
family of cuprate superconductors contains no copper-oxide chains, and is nearly isotropic in the
square-lattice copper-oxide plane. In this case, one is therefore expect that the experiments in
magnetic field reflect the intrinsic aspects of the square-lattice copper-oxide plane response
\cite{Fradkin15}. However, the experimental data of the nematic-order state strength dependence
of the electromagnetic response in cuprate superconductors are still lacking to date, i.e., it is
still unclear how the intrinsic features of the electromagnetic response evolves with the strength
of the electronic nematicity. Furthermore, to the best of our knowledge, the intrinsic features of
the electromagnetic response of cuprate superconductors with coexisting electronic nematicity have
also not been discussed starting from a SC theory so far. In this case, the crucial issue is to
understand the exotic properties of the electromagnetic response in cuprate superconductors with
coexisting electronic nematicity even from a theoretical analysis.

In despite of the experimental developments
\cite{Zhang94,Basov95,Sun95,Barnea04,Kiefl10,Zhang18,Razzoli17,Quijada99}, the role of the
nematic order, such as whether it favor or compete with superconductivity and how it relates to
the spontaneous translation symmetry breaking of the electromagnetic response, still remains
controversial. Some numerical analyses show that the nematic order competes possibly with the
electron pairing \cite{Plakida00,Edegger06,Miyanaga06,Wollny09}. On the other hand, the
interesting theoretical idea of the nematic-order-driven superconductivity has been put forward,
where the fluctuations associated with the electronically nematic order can enhance
superconductivity \cite{Kitatani17,Maier14,Lederer15,Kaczmarczyk16,Lederer17,Kao05,Lee16}.
In our recent works \cite{Cao22,Cao21}, the electronic structure of cuprate superconductors with
coexisting electronic nematicity has been investigated based on the kinetic-energy-driven
superconductivity, where we have shown that superconductivity is enhanced by the electronic
nematicity. In particular, we \cite{Cao21} have also shown that the characteristic energy of the
nematic-order state as a function of the nematic-order state strength presents a similar behavior
of $T_{\rm c}$, which therefore suggests a possible connection between the characteristic energy
of the nematic-order state and the enhancement of the superconductivity. In this paper, we study
the nematic-order state strength dependence of the {\it in-plane} electromagnetic response in
cuprate superconductors along with this line. Our results indicate that the electromagnetic
response of cuprate superconductors with coexisting electronic nematicity is inequivalent along
with the $\hat{a}$- and $\hat{b}$-axes. In particular, we show that in addition to the pure
d-wave component of the SC gap, the pure s-wave component of the SC gap is generated by the
electronic nematicity, therefore there is a coexistence and competition between the pure d-wave
and s-wave components. Moreover, this coexistence and competition leads to the SC condensation
energy that first increases with the strength of the electronic nematicity in the weak strength
region, then reaches a maximum value at around the optimal strength of the electronic nematicity,
but is suppressed with further increase of the strength in the strong strength region of the
electronic nematicity. This dome-like shape of the nematic-order state strength dependence of the
SC condensation energy therefore in turn induces the enhancement of superconductivity, and gives
rise to the dome-like shape of the nematic-order state strength dependence of the superfluid
density.

The organization of the paper is as follows. In Sec. \ref{general-framework}, the response
kernel with broken rotation symmetry is derived  based on the linear response approach for a
purely transverse vector potential, and then this response kernel is employed to discuss the
rotation symmetry-breaking of the Meissner effect of cuprate superconductors with coexisting
electronic nematicity in the long wavelength limit in Sec. \ref{meissner-effect}, where the
local magnetic-field profiles along the $\hat{a}$- and $\hat{b}$-axes are derived based on the
specular reflection model, and results show that the distance dependence of the local
magnetic-field profile follows an exponential law as was expected for the local electrodynamic
response. Finally, a summary is given in Sec. \ref{conclusions}. In the Appendix, we presents
the derivation of the electron propagator by taking into account the vertex correction.

\section{Formalism of electromagnetic response with broken rotation symmetry}
\label{general-framework}

\subsection{Model and propagator}\label{model}

As we have mentioned in Sec. \ref{Introduction}, in what concerns the electronic properties of
cuprate superconductors, it is widely accepted that the motion of the electrons which play the
crucial role for superconductivity are restricted to the square-lattice copper-oxide plane
\cite{Cooper94,Takenaka94}. Soon after the discovery of superconductivity in cuprate
superconductors, it has been proposed that the essential physics of the doped copper-oxide
plane can be captured by the $t$-$J$ model on a square-lattice \cite{Anderson87}. However, for
discussions of the electromagnetic response of cuprate superconductors with coexisting electronic
nematicity, the coupling between an external magnetic-field and the electrons can be treated via
the Peierls construction, in which the electron creation and annihilation operators develop a
phase factor, and then the resulting $t$-$J$ model is obtained as \cite{Liu20},
\begin{eqnarray}\label{tJ-model}
H&=&-\sum_{l\hat{\eta}\sigma}t_{\hat{\eta}}e^{-i({e}/{\hbar}){\bf A}(l)\cdot\hat{\eta}}
C^{\dagger}_{l\sigma}C_{l+\hat{\eta}\sigma}\nonumber\\
&+&\sum_{l\hat{\eta}'\sigma}t'_{\hat{\eta}'}e^{-i({e}/{\hbar}){\bf A}(l)\cdot\hat{\eta}'}
C^{\dagger}_{l\sigma}C_{l+\hat{\eta}'\sigma}
\nonumber\\
&+&\mu\sum_{l\sigma}C^{\dagger}_{l\sigma}C_{l\sigma} +\sum_{l\hat{\eta}}J_{\hat{\eta}}
{\bf S}_{l}\cdot {\bf S}_{l+\hat{\eta}},
\end{eqnarray}
where $t_{\hat{\eta}}$ and $J_{\hat{\eta}}$ are the hoping amplitude and exchange coupling,
respectively, for the nearest neighbors $\hat{\eta}$, $t'_{\hat{\eta}'}$ is the hoping amplitude
for the next nearest neighbors $\hat{\eta}'$, $C^{\dagger}_{l\sigma}$ ($C_{l\sigma}$) creates
(annihilates) an electron at site $l$ with spin $\sigma$, ${\bf S}_{l}$ is the spin operator with
its components $S^{\rm x}_{l}$, $S^{\rm y}_{l}$, and $S^{\rm z}_{l}$, and $\mu$ is the chemical
potential. Following the previous analyses of the exotic features of cuprate superconductors with
coexisting electronic nematicity
\cite{Plakida00,Edegger06,Miyanaga06,Wollny09,Kitatani17,Maier14,Lederer15,Kaczmarczyk16,Lederer17,Kao05,Lee16,Cao22,Cao21},
the next nearest-neighbor (NN) hoping amplitude in the $t$-$J$ model (\ref{tJ-model}) is chosen as
$t'_{\hat{\tau}}=t'$, while the NN hoping amplitude $t_{\hat{\eta}}$ can be chosen as,
\begin{eqnarray}
t_{\hat{x}}&=&(1-\varsigma)t, ~~~~ t_{\hat{y}}=(1+\varsigma)t.  \label{NN-hoping}
\end{eqnarray}
This NN hoping amplitude in Eq. (\ref{NN-hoping}) is strongly anisotropic along the ${\hat{a}}$
and ${\hat{b}}$ axes, and can give a consistent description of the ARPES spectrum in the
nematic-order state within the standard tight-binding model \cite{Nakata18}. Concomitantly, this
anisotropic NN hoping amplitude in Eq. (\ref{NN-hoping}) induces the anisotropic NN exchange
coupling $J_{\hat{x}}=(1-\varsigma)^{2}J$ and $J_{\hat{y}}=(1+\varsigma)^{2}J$ in the $t$-$J$
model (\ref{tJ-model}). Moreover, this anisotropic parameter $\varsigma$ in Eq. (\ref{NN-hoping})
depicts the orthorhombicity of the electronic structure, and this is why it has been identified
as the strength of the electronic nematicity in the system \cite{Nakata18}. On the other hand,
this anisotropic parameter $\varsigma$ can be also thought to be a variational parameter, and
then in a given doping concentration, this anisotropic parameter $\varsigma$ can be determined
self-consistently by minimizing the energy as the discussions based on the variational Monte
Carlo approach \cite{Edegger06}. In this paper, this anisotropic parameter $\varsigma$ in the
SC-state is determined self-consistently by maximizing the SC condensation energy (then
lowering the total energy) as shown in Fig. \ref{energy-nematicity}. Although the nematic-order
state strength dependence of the electromagnetic response is discussed, the calculated results
of the magnetic-field penetration-depths along the $\hat{a}$- and $\hat{b}$-axes in Sec.
\ref{meissner-effect} only for the optimal strength of the electronic nematicity, where the
lowest total energy is achieved, are used to compare with the corresponding experimental data
\cite{Kiefl10,Quijada99}. The anisotropic NN
hoping amplitudes in Eq. (\ref{NN-hoping}) also indicate that the rotation symmetry is broken
already in the starting $t$-$J$ model (\ref{tJ-model}). Throughout this paper, we choose the
parameters $t/J=3$ and $t'/t=1/3$ as in our previous discussions \cite{Cao22,Cao21}. However,
when necessary to compare with the experimental data, we take $J=100$ meV.

The strong electron correlation in the $t$-$J$ model (\ref{tJ-model}) manifests itself by the
restriction of the motion of the electrons in a Hilbert subspace without a double electron
occupancy \cite{Feng93,Zhang93,Yu92,Lee06,Edegger07}, i.e.,
$\sum_{\sigma}C^{\dagger}_{l\sigma}C_{l\sigma}\leq 1$, which can be treated properly in
analytical calculations in terms of the fermion-spin transformation \cite{Feng0494,Feng15}, in
which the constrained electron operators $C_{l\uparrow}$ and $C_{l\downarrow}$ are replaced by,
\begin{eqnarray}\label{CSS}
C_{l\uparrow}=h^{\dagger}_{l\uparrow}S^{-}_{l}, ~~~~
C_{l\downarrow}=h^{\dagger}_{l\downarrow}S^{+}_{l},
\end{eqnarray}
respectively, where the spinful fermion operator $h_{l\sigma}=e^{-i\Phi_{l\sigma}}h_{l}$ keeps
track of the charge degree of freedom of the constrained electron together with some effects of
spin configuration rearrangements due to the presence of the doped hole itself (charge carrier),
while the spin operator $S_{l}$ keeps track of the spin degree of freedom of the constrained
electron. The main advantage of this fermion-spin approach (\ref{CSS}) is that the on-site local
constraint of no double electron occupancy is satisfied in actual calculations.

Superconductivity in cuprate superconductors arises from the binding of electrons into electron
pairs, thereby forming a superfluid with a SC gap in the single-particle excitation spectrum.
Although it is believed that the electron pairing is due to the coupling of the electrons to
particular bosonic excitations, the nature of these bosonic excitations remains controversial,
where two main proposals are disputing the explanations of the bosonic glue to hold the electron
pairs together. In one of the proposals, the electron pairing is associated to the phonon
\cite{Iwasawa08,Zhou07,Rosch04,Rosch04-1,Lanzara01}, while in the other, the electron pairing
is related to the spin excitation \cite{Monthoux91,Monthoux07,Feng0306,Feng12,Feng15a}.
In the case of zero magnetic-field, the kinetic-energy-driven SC mechanism has been developed
\cite{Feng15,Feng0306,Feng12,Feng15a} based on the $t$-$J$ model (\ref{tJ-model}) in the
fermion-spin representation (\ref{CSS}), where the charge carriers are held together in the
d-wave pairs at low temperatures by the attractive interaction that originates directly from the
kinetic energy of the $t$-$J$ model by the exchange of a strongly dispersive spin excitation,
then the d-wave electron pairs originated from the d-wave charge-carrier pairs are due to the
charge-spin recombination, and these d-wave electron pairs condense to the SC-state with the
d-wave symmetry. The characteristic features of the kinetic-energy-driven SC mechanism can be
also summarized as: (i) the mechanism is purely electronic without phonons; (ii) the mechanism
indicates that the strong electron correlation favors superconductivity, since the main
ingredient is identified into an electron pairing mechanism not involving the phonon, the
external degree of freedom, but the internal spin degree of freedom of the constrained
electron; (iii) the SC-state is controlled by both the SC gap and quasiparticle coherence,
leading to that the maximal $T_{\rm c}$ occurs around the optimal doping, and then decreases
in both the underdoped and the overdoped regimes. Very recently, the framework of this
kinetic-energy-driven
superconductivity has been generalized to discuss the intertwinement of superconductivity with
the electronic nematicity \cite{Cao22,Cao21}, where the breaking of the rotation symmetry due to
the presence of the electronic nematicity is verified by the inequivalence on the average of the
electronic structure at the two Bragg scattering sites. However, in the above discussions
\cite{Cao22,Cao21}, the vertex correction for the electron self-energy has been ignored. In the
following discussions, we study the exotic features of the electromagnetic response of cuprate
superconductors with coexisting electronic nematicity by taking into account the vertex correction
for the electron self-energy. Following these recent discussions at zero magnetic-field
\cite{Cao22,Cao21}, the vertex corrected electron propagator of the $t$-$J$ model (\ref{tJ-model})
in the SC-state with coexisting nematic order can be obtained explicitly in the Nambu
representation as [see Appendix \ref{Green-function}],
\begin{equation}\label{NPEGF}
\mathbb{G}_\varsigma({\bf k},\omega)=Z^{(\varsigma)}_{\rm F}{\omega\tau_{0}
+\bar{\varepsilon}^{(\varsigma)}_{\bf k}\tau_{3}-\bar{\Delta}^{(\varsigma)}_{\rm Z}({\bf k})
\tau_{1}\over\omega^2-E^{(\varsigma)2}_{\bf k}}
\end{equation}
where $\tau_{0}$ is a unit matrix, $\tau_{1}$ and $\tau_{3}$ are Pauli matrices,
$E^{(\varsigma)}_{\bf k} = \sqrt {\bar{\varepsilon}^{(\varsigma)2}_{\bf k}
+|\bar{\Delta}^{(\varsigma)}_{\rm Z}({\bf k})|^{2}}$ is the SC quasiparticle energy spectrum,
$\bar{\varepsilon}^{(\varsigma)}_{\bf k}=Z^{(\varsigma)}_{\rm F}\varepsilon^{(\varsigma)}_{\bf k}$
is the renormalized electron orthorhombic energy dispersion,
$\bar{\Delta}^{(\varsigma)}_{\rm Z}({\bf k})=Z^{(\varsigma)}_{\rm F}
\bar{\Delta}^{(\varsigma)}({\bf k})$ is the renormalized SC gap,
$\varepsilon^{(\varsigma)}_{\bf k}$ is the orthorhombic energy dispersion in the tight-binding
approximation, and has been obtained directly from the $t$-$J$ model (\ref{tJ-model}) as
\cite{Cao22,Cao21},
\begin{eqnarray}\label{band-structure}
\varepsilon^{(\varsigma)}_{\bf k}=-4t[(1-\varsigma)\gamma_{{\bf k}_{x}}+(1+\varsigma)
\gamma_{{\bf k}_{y}}]+4t'\gamma_{\bf k}'+\mu,
\end{eqnarray}
with $\gamma_{{\bf k}_{x}}={\rm cos}k_{x}/2$, $\gamma_{{\bf k}_{y}}={\rm cos}k_{y}/2$,
$\gamma_{\bf k}'={\rm cos}k_{x}{\rm cos}k_{y}$, and $\bar{\Delta}^{(\varsigma)}({\bf k})$ is
the SC gap, and can be expressed as,
\begin{eqnarray}
\bar{\Delta}^{(\varsigma)}({\bf k})=\bar{\Delta}^{(\varsigma)}_{\hat{x}}\gamma_{{\bf k}_{x}}
-\bar{\Delta}^{(\varsigma)}_{\hat{y}}\gamma_{{\bf k}_{y}}, \label{SCGF}
\end{eqnarray}
while the quasiparticle coherent weight $Z^{(\varsigma)}_{\rm F}$ and the components of the SC
gap parameter $\bar{\Delta}^{(\varsigma)}_{\hat{x}}$ and $\bar{\Delta}^{(\varsigma)}_{\hat{y}}$
are given in Appendix \ref{Green-function}.

In particular, the SC gap in Eq. (\ref{SCGF}) can be also rewritten explicitly as,
\begin{eqnarray}
\bar{\Delta}^{(\varsigma)}({\bf k})=\bar{\Delta}^{(\varsigma)}_{\rm d}\gamma^{\rm (d)}_{\bf k}
+\bar{\Delta}^{(\varsigma)}_{\rm s}\gamma^{\rm (s)}_{\bf k}, \label{SCGF-DS}
\end{eqnarray}
where $\gamma^{\rm (d)}_{\bf k}=({\rm cos}k_{x}-{\rm cos}k_{y})/2$,
$\gamma^{\rm (s)}_{\bf k}=({\rm cos}k_{x}+{\rm cos}k_{y})/2$, and
$\bar{\Delta}^{(\varsigma)}_{\rm d}=(\bar{\Delta}^{(\varsigma)}_{\hat{x}}
+\bar{\Delta}^{(\varsigma)}_{\hat{y}})/2$ and $\bar{\Delta}^{(\varsigma)}_{\rm s}
=(\bar{\Delta}^{(\varsigma)}_{\hat{x}}-\bar{\Delta}^{(\varsigma)}_{\hat{y}})/2$ are the d-wave
and s-wave components of the SC gap parameter, respectively. The above result in
Eq. (\ref{SCGF-DS}) therefore show that the symmetry of the SC-state with coexisting nematic
order is modified from the pure d-wave electron pairing to the d+s wave \cite{Edegger06}. In
other words, unlike the electronic structure with the four-fold rotation symmetry, the
electronic structure with the two-fold rotation symmetry can not have a pure d-wave SC gap.
Moreover, it should be noted that the d-wave component of the SC gap parameter
$\bar{\Delta}^{(\varsigma)}_{\rm d}$ is also equal to the maximal SC gap parameter at $[0,\pi]$
point of the Brillouin zone (then at around the antinodal regime). Moreover, it should be
emphasized that although the result in Eq. (\ref{NPEGF}) is the basic Bardeen-Cooper-Schrieffer
formalism \cite{Matsui03}, the electron pairing mechanism is driven by the kinetic energy by
the exchange of a strongly dispersive spin excitation \cite{Feng15,Feng0306,Feng12,Feng15a}. In
particular, based on this result in Eq. (\ref{NPEGF}), the evolutions of $T_{\rm c}$ with the
doping concentration and nematic-order state strength have been investigated in terms of the
self-consistent calculations at the condition of the SC gap $\bar{\Delta}^{(\varsigma)}=0$,
where the main results can be summarized as: (i) in the case of the absence of the nematic
order, $T_{\rm c}$ obtained in our previous works in Refs. \onlinecite{Feng12,Feng15a} has a dome-like
shape doping dependence with the maximum $T_{\rm c}$ that occurs at around the optimal doping
$\delta\sim 0.15$, in good agreement with the corresponding experimental results observed on
cuprate superconductors \cite{Drozdov18}; (ii) for an any given doping, $T_{\rm c}$ obtained in
our recent works in Refs. \onlinecite{Cao22,Cao21} increases with the increase of the strength of the
electronic nematicity in the weak strength region, and reaches a maximum in the optimal strength,
then decreases with the increase of the strength of the electronic nematicity in the strong
strength region. This dome-like shape nematic-order state strength dependence of $T_{\rm c}$
thus indicates that the electronic nematicity enhances superconductivity \cite{Cao22,Cao21}. In
particular, it should be emphasized that these results of the doping dependence of $T_{\rm c}$
obtained in our previous works in Refs. \onlinecite{Feng12,Feng15a} and nematic-order state strength
dependence of $T_{\rm c}$ obtained in our recent works in Refs. \onlinecite{Cao22,Cao21} are evaluated
by the self-consistent calculation without using any adjustable parameters, and in this sense,
our calculations for the doping and nematic-order state strength dependence of $T_{\rm c}$ are
controllable.

\subsection{Rotation symmetry-breaking of response kernel}\label{kernel-function}

The weak external magnetic-field applied to the system usually represents a weak perturbation,
however, the induced field generated by supercurrents can cancel this weak external
magnetic-field over most of the volume of the sample. Concomitantly, the net field acts only very
near the surface on a scale of the magnetic-field penetration depth and so it can be treated as a
weak perturbation on the system as a whole. In this case, the electromagnetic response can be
successfully studied based on the linear response approach \cite{Fukuyama69,Misawa94,Kostyrko94},
where the electron current density ${\bf J}^{(\varsigma)}$ of the induced microscopic screening
current and the vector potential ${\bf A}$ satisfies the general relation,
\begin{equation}\label{linres}
J^{(\varsigma)}_{\mu}({\bf q},\omega)=-\sum\limits_{\nu=1}^{3}
K_{\mu\nu}(\varsigma,{\bf q},\omega)A_{\nu}({\bf q},\omega),
\end{equation}
with the Greek indices that label the axes of the Cartesian coordinate system, while
$K_{\mu\nu}(\varsigma,{\bf q},\omega)$ is a nonlocal response kernel. The way the system reacts to
a weak electromagnetic stimulus is entirely described by this response kernel. Once this response
kernel is known, the effect of a weak external magnetic field can be quantitatively characterized
by experimentally measurable quantities. This response kernel (\ref{linres}) can be broken up into
its diamagnetic and paramagnetic parts as,
\begin{equation}\label{kernel}
K_{\mu\nu}(\varsigma,{\bf q},\omega)=K^{({\rm d})}_{\mu\nu}(\varsigma,{\bf q},\omega)
+K^{({\rm p})}_{\mu\nu}(\varsigma,{\bf q},\omega).
\end{equation}

In the general relation between the electron current density ${\bf J}^{(\varsigma)}$ and the
vector potential ${\bf A}$ in Eq. (\ref{linres}), the vector potential ${\bf A}$ is coupled to
the electrons, which are now represented by the electron operators in the fermion-spin
transformation (\ref{CSS}). For the evaluation of the electron current density, it is needed to
obtain the electron polarization operator, which is defined as a summation over all the particles
and their positions, and can be calculated straightforwardly in terms of the fermion-spin
transformation (\ref{CSS}) as,
\begin{equation}\label{poloper}
{\bf P}=-e\sum\limits_{l\sigma}{\bf R}_{l}C^{\dagger}_{l\sigma}C_{l\sigma}
=e\sum\limits_{l}{\bf R}_{l}h^{\dagger}_{l} h_{l},
\end{equation}
then the electron current density is obtained by the calculation of the time-derivative of this
electron polarization operator (\ref{poloper}) as \cite{Liu20},
\begin{eqnarray}\label{current-density-operator}
{\bf J}_{\varsigma}&=&{\partial {\bf P}\over\partial t}={i\over\hbar}[H,{\bf P}]\nonumber\\
&=& {ie\over\hbar}\sum\limits_{l{\hat{\eta}}\sigma}t_{\hat{\eta}}{\hat{\eta}}
e^{-i{e\over\hbar}{\bf A}(l)\cdot{\hat{\eta}}}C^{\dagger}_{l\sigma} C_{l+\hat{\eta}\sigma}
\nonumber\\
&+& {ie\over\hbar}\sum\limits_{l{\hat{\eta}'}\sigma}t_{\hat{\eta}'}{\hat{\eta}'}
e^{-i{e\over\hbar}{\bf A}(l)\cdot{\hat{\eta}'}}C^{\dagger}_{l\sigma} C_{l+\hat{\eta}'\sigma}.
~~~~~~
\end{eqnarray}
In corresponding to the diamagnetic and paramagnetic parts of the response kernel in
Eq. (\ref{kernel}), we can express this electron current density in
Eq. (\ref{current-density-operator}) as
${\bf J}_{\varsigma}={\bf J}^{({\rm d})}_{\varsigma}+{\bf J}^{({\rm p})}_{\varsigma}$, while the
diamagnetic and paramagnetic parts ${\bf J}^{({\rm d})}_{\varsigma}$ and
${\bf J}^{({\rm p})}_{\varsigma}$ are obtained in the linear response theory as \cite{Liu20},
\begin{subequations}
\begin{eqnarray}
{\bf J}^{({\rm d})}_{\varsigma}&=&{e^{2}\over\hbar^{2}}\sum\limits_{l{\hat{\eta}}\sigma}
t_{\hat{\eta}}{\hat{\eta}}{\bf A}(l)\cdot{\hat{\eta}}C^{\dagger}_{l\sigma} C_{l+\hat{\eta}\sigma}
\nonumber\\
&-&{e^{2}\over\hbar^{2}}t'\sum\limits_{l{\hat{\eta}'}\sigma}{\hat{\eta}'}{\bf A}(l)
\cdot{\hat{\eta}'}C^{\dagger}_{l\sigma} C_{l+\hat{\eta'}\sigma},\label{tcurdia}\\
{\bf J}^{({\rm p})}_{\varsigma}&=&{ie\over\hbar}\sum\limits_{l{\hat{\eta}}\sigma}t_{\hat{\eta}}
{\hat{\eta}}C^{\dagger}_{l\sigma}C_{l+\hat{\eta}\sigma}-{ie\over\hbar}
t'\sum\limits_{l{\hat{\eta}'}\sigma}{\hat{\eta}'}C^{\dagger}_{l\sigma}C_{l+\hat{\eta}'\sigma},
~~~~~~\label{tcurpara9}
\end{eqnarray}
\end{subequations}
respectively. The above result in Eq. (\ref{tcurdia}) shows that the diamagnetic current is
directly proportional to the vector potential. In this case, it is thus straightforward to obtain
the diamagnetic part of the response kernel as,
\begin{subequations}\label{diakernel}
\begin{eqnarray}
K_{\hat{x}\hat{x}}^{({\rm d})}(\varsigma,{\bf q},\omega)&=&{4e^{2}\over\hbar^{2}}
[\phi_{{\rm c1}\hat{x}}(1-\varsigma)t-2\phi_{\rm c2}t']\nonumber\\
&=&{1\over\mu_{0}\lambda^{2}_{La}(\varsigma,T)},~~~~~~~\label{diakernel-x}\\
K_{\hat{y}\hat{y}}^{({\rm d})}(\varsigma,{\bf q},\omega)&=&{4e^{2}\over\hbar^{2}}
[\phi_{{\rm c1}\hat{y}}(1+\varsigma)t-2\phi_{\rm c2}t']\nonumber\\
&=& {1\over\mu_{0}\lambda^{2}_{Lb}(\varsigma,T)},\label{diakernel-y}\\
K_{\hat{x}\hat{y}}^{({\rm d})}(\varsigma,{\bf q},\omega)
&=&K_{\hat{y}\hat{x}}^{({\rm d})}(\varsigma,{\bf q},\omega)=0,
\end{eqnarray}
\end{subequations}
where $\mu_{0}$ is the magnetic permeability, and $\lambda_{La}(\varsigma,T)$ and
$\lambda_{Lb}(\varsigma,T)$ are the London penetration depths along the $\hat{a}$- and
$\hat{b}$-axes, respectively, while the electron particle-hole parameters
$\phi_{{\rm c1}\hat{x}}=\langle C^{\dagger}_{l\sigma} C_{l+\hat{x}\sigma}\rangle$,
$\phi_{{\rm c1}\hat{y}}=\langle C^{\dagger}_{l\sigma} C_{l+\hat{y}\sigma}\rangle$,
and $\phi_{\rm c2}=\langle C^{\dagger}_{l\sigma}C_{l+\hat{\eta}'\sigma}\rangle$ are
evaluated directly from the electron diagonal propagator (\ref{NPEGF}) as,
\begin{subequations}\label{particle-hole-parameters}
\begin{eqnarray}
\phi_{{\rm c1}\hat{x}}&=&{1\over 2N}\sum_{{\bf k}}\gamma_{{\bf k}_{x}}Z^{(\varsigma)}_{\rm F}
\left (1-{\bar{\varepsilon}^{(\varsigma)}_{{\bf k}}\over E^{(\varsigma)}_{\bf k}}{\rm tanh}
[{1\over 2}\beta E^{(\varsigma)}_{\bf k}]\right ), ~~~~~~~~\\
\phi_{{\rm c1}\hat{y}}&=&{1\over 2N}\sum_{{\bf k}}\gamma_{{\bf k}_{y}}Z^{(\varsigma)}_{\rm F}
\left (1-{\bar{\varepsilon}^{(\varsigma)}_{{\bf k}}\over E^{(\varsigma)}_{\bf k}}{\rm tanh}
[{1\over 2}\beta E^{(\varsigma)}_{\bf k}]\right ),\\
\phi_{\rm c2}&=&{1\over 2N}\sum_{{\bf k}}\gamma_{{\bf k}}'Z^{(\varsigma)}_{\rm F}
\left (1-{\bar{\varepsilon}^{(\varsigma)}_{{\bf k}}\over E^{(\varsigma)}_{\bf k}}{\rm tanh}
[{1\over 2}\beta E^{(\varsigma)}_{\bf k}]\right ),
\end{eqnarray}
\end{subequations}
with the number of sites on a square lattice $N$.

However, the derivation of the paramagnetic part of the response kernel is rather complicated,
since it can be obtained as
$K_{\mu\nu}^{({\rm p})}(\varsigma,{\bf q},\omega)=P^{(\varsigma)}_{\mu\nu}({\bf q},\omega)$,
with electron current-current correlation function \cite{Fukuyama69,Misawa94,Kostyrko94},
\begin{equation}\label{corP}
P^{(\varsigma)}_{\mu\nu}({\bf q},\tau)=-\langle T_{\tau}
J^{({\rm p})}_{\mu}(\varsigma,{\bf q},\tau)J_{\nu}^{({\rm p})}(\varsigma,-{\bf q},0)\rangle .
\end{equation}
If the gauge invariant is kept in the theory, it is crucial to derive properly the above
electron correlation function (\ref{corP}) in a way maintaining local charge conservation
\cite{Schrieffer64,Fukuyama69,Misawa94,Kostyrko94}. In the following calculations, we work
with a fixed gauge of the vector potential as in the previous discussions \cite{Liu20}. For
a convenience in the calculation of the above electron current-current correlation
function (\ref{corP}), the electron operators can be rewritten in the Nambu representation
as $\Psi^{\dagger}_{\bf k}=(C^{\dagger}_{{\bf k}\uparrow},C_{-{\bf k}\downarrow})$ and
$\Psi_{{\bf k}+{\bf q}}=(C_{{\bf k}+{\bf q}\uparrow},C^{\dagger}_{-{\bf k}-{\bf q}
\downarrow})^{\rm T}$. In this case, the electron density is summed over the position of all
electrons, and then its Fourier transform in the Nambu notation can be expressed as
$\rho({\bf q})=(e/N)\sum_{{\bf k}}\Psi^{\dagger}_{\bf k}\tau_{3}\Psi_{{\bf k}+{\bf q}}$.
According to this expression of the electron density and the paramagnetic part of the electron
current density in Eq. (\ref{tcurpara9}), we now can express the paramagnetic four-current
density in the Nambu representation as,
\begin{eqnarray}\label{curnambu}
J_{\mu}^{({\rm p})}(\varsigma,{\bf q})={1\over N}\sum\limits_{\bf k}\Psi^{\dagger}_{{\bf k}}
{\mathbf{\gamma}}^{(\varsigma)}_{\mu}({\bf k},{\bf q})\Psi_{{\bf k}+{\bf q}},
\end{eqnarray}
with the bare current vertex,
\begin{eqnarray}
{\mathbf{\gamma}}^{(\varsigma)}_{\mu}({\bf k,q})= \left \{
\begin{array}{lr}
-\frac{2e}{\hbar}e^{\frac{1}{2}iq_\mu}\{{\rm sin}(k_\mu+\frac{1}{2}q_\mu)\\
\times[t_{\mu}-2t'{\sum\limits_{\nu\ne\mu}}{\rm cos}(\frac{1}{2}q_\nu)
{\rm cos}(k_\nu+\frac{1}{2}q_\nu)]
\\
-i(2t'){\rm cos}(k_\mu+\frac{1}{2}q_\mu)\\
\times\sum\limits_{\nu\ne\mu}{\rm sin}(\frac{1}{2}q_\nu){\rm sin}(k_\nu+\frac{1}{2}q_\nu)\}
\tau_{0},{\rm for}\ \mu\ne0\\
e\tau_{3},  {\rm for}\ \mu=0
\end{array}\right.\label{barevertex}
\end{eqnarray}
It should be emphasized that we are calculating the electron current-current correlation function
(\ref{corP}) with the paramagnetic current density operator (\ref{curnambu}), i.e., bare current
vertex (\ref{barevertex}), but the electron propagator (\ref{NPEGF}). Concomitantly, we do not
take into account longitudinal excitations properly in this scenario \cite{Schrieffer64}, and the
obtained results are valid only in the gauge, where the vector potential is purely transverse, e.g.
in the Coulomb gauge. In this case, the electron current-current correlation function (\ref{corP})
in the Nambu representation can be expressed in terms of the electron propagator (\ref{NPEGF}) as,
\begin{eqnarray}
&&P^{(\varsigma)}_{\mu\nu}({\bf q},iq_{m})={1\over N}\sum\limits_{\bf k}
{\mathbf{\gamma}}^{(\varsigma)}_{\mu}({\bf k},{\bf q})
{\mathbf{\gamma}}^{(\varsigma)*}_{\nu}({\bf k},{\bf q})\nonumber\\
&\times&{1\over\beta}\sum\limits_{i\omega_{n}}{\rm{Tr}}\left[{\mathbb{G}_{\varsigma}}({\bf k}
+{\bf q},i\omega_{n}+iq_{m}){\mathbb{G}_{\varsigma}}({\bf k},i\omega_{n})\right],
\label{barepolmats}
\end{eqnarray}
where $\omega_{n}$ and $q_{m}$ are the fermionic and bosonic Matsubara frequencies, respectively.
Substituting the electron propagator (\ref{NPEGF}) into the above Eq. (\ref{barepolmats}),
and performing the summation over fermionic Matsubara frequencies, the paramagnetic part of the
response kernel $K_{\mu\nu}^{({\rm p})}(\varsigma,{\bf q},\omega)$ in the static limit
($\omega\sim 0$) can be derived as,
\begin{subequations}\label{parakernel}
\begin{eqnarray}
K_{\hat{x}\hat{x}}^{({\rm p})}(\varsigma,{\bf q},0)&=&{1\over N}\sum_{\bf k}
\mathbf{\gamma}^{(\varsigma)}_{x}({\bf k},{\bf q})
\mathbf{\gamma}^{(\varsigma)*}_{x}({\bf k},{\bf q})\nonumber\\
&\times&[L^{(\varsigma)}_{\rm c1}({\bf k},{\bf q})+L^{(\varsigma)}_{\rm c2}({\bf k},{\bf q})],\\
K_{\hat{y}\hat{y}}^{({\rm p})}(\varsigma,{\bf q},0)&=&{1\over N}\sum_{\bf k}
\mathbf{\gamma}^{(\varsigma)}_{y}({\bf k},{\bf q})
\mathbf{\gamma}^{(\varsigma)*}_{y}({\bf k},{\bf q})\nonumber\\
&\times&[L^{(\varsigma)}_{\rm c1}({\bf k},{\bf q})+L^{(\varsigma)}_{\rm c2}({\bf k},{\bf q})],
\end{eqnarray}
\begin{eqnarray}
K_{\hat{x}\hat{y}}^{({\rm p})}(\varsigma,{\bf q},0)&=&
K_{\hat{y}\hat{x}}^{({\rm p})}(\varsigma,{\bf q},0)=0,
\end{eqnarray}
\end{subequations}
where the key functions $L^{(\varsigma)}_{\rm c1}({\bf k},{\bf q})$ and
$L^{(\varsigma)}_{\rm c2}({\bf k},{\bf q})$ are given by,
\begin{subequations}
\begin{eqnarray}
L^{(\varsigma)}_{\rm c1}({\bf k},{\bf q}) &=& Z_{\rm F}^{(\varsigma)2}\left [
1+{\bar{\varepsilon}^{(\varsigma)}_{{\bf k}+{\bf q}}
\bar{\varepsilon}^{(\varsigma)}_{\bf k}+\bar{\Delta}^{(\varsigma)}_{\rm Z}({\bf k}+{\bf q})
\bar{\Delta}^{(\varsigma)}_{\rm Z}({\bf k})\over E^{(\varsigma)}_{\bf k}
E^{(\varsigma)}_{{\bf k}+{\bf q}}}\right ]\nonumber\\
&\times&{n_{\rm F}(E^{(\varsigma)}_{\bf k})-n_{\rm F}(E^{(\varsigma)}_{{\bf k}+{\bf q}})
\over E^{(\varsigma)}_{\bf k}-E^{(\varsigma)}_{{\bf k} +{\bf q}}}, ~~~\label{key-function-1}\\
L^{(\varsigma)}_{\rm c2}({\bf k},{\bf q}) &=& Z_{\rm F}^{(\varsigma)2}\left [
1-{\bar{\varepsilon}^{(\varsigma)}_{{\bf k}+{\bf q}}
\bar{\varepsilon}^{(\varsigma)}_{\bf k}+\bar{\Delta}^{(\varsigma)}_{\rm Z}({\bf k}+{\bf q})
\bar{\Delta}^{(\varsigma)}_{\rm Z}({\bf k})\over E^{(\varsigma)}_{\bf k}
E^{(\varsigma)}_{{\bf k}+{\bf q}}}\right ]\nonumber\\
&\times&{n_{\rm F}(E^{(\varsigma)}_{\bf k})+n_{\rm F}(E^{(\varsigma)}_{{\bf k}+{\bf q}})
-1\over E^{(\varsigma)}_{\bf k}+E^{(\varsigma)}_{{\bf k}+{\bf q}}}.~~~~~~~~\label{key-function-2}
\end{eqnarray}
\end{subequations}
Incorporating these results of the paramagnetic part of the response kernel (\ref{parakernel})
with the corresponding results of the diamagnetic part of the response kernel (\ref{diakernel}),
the response kernels in the presence of the electronic nematicity along the $\hat{a}$- and
$\hat{b}$-axes now are obtained as,
\begin{subequations}\label{kernel-5}
\begin{eqnarray}
K_{\hat{x}\hat{x}}(\varsigma,{\bf q},0)&=&{1\over\mu_{0}\lambda^{2}_{La}(\varsigma,T)}
+K_{\hat{x}\hat{x}}^{({\rm p})}(\varsigma,{\bf q},0), \label{kernel-5-x}\\
K_{\hat{y}\hat{y}}(\varsigma,{\bf q},0)&=&{1\over\mu_{0}\lambda^{2}_{Lb}(\varsigma,T)}
+K_{\hat{y}\hat{y}}^{({\rm p})}(\varsigma,{\bf q},0), \label{kernel-5-y}
\end{eqnarray}
\end{subequations}
respectively.



\section{Quantitative characteristics of rotation symmetry-breaking of electromagnetic response
in the long wavelength limit} \label{meissner-effect}

In this Section, we discuss the electromagnetic response of cuprate superconductors with
coexisting electronic nematicity in the long wavelength limit. As in the previous discussions
\cite{Liu20}, we introduce a characteristic-length scale $a_{0}=\sqrt{\hbar^{2}a/\mu_{0}e^{2}J}$
for a convenience in the following discussions, where $a$ is the lattice constant. Using the
lattice constant $a\approx 0.383$ nm of YBa$_2$Cu$_3$O$_{7-y}$, this characteristic-length is
obtained as $a_{0}\approx 97.8$ nm.

\subsection{Nematic-order state strength dependence of superfluid density}
\label{average-superfluid-density}

In the long wavelength limit $|{\bf q}|\to 0$, the function
$L^{(\varsigma)}_{c2}({\bf k,q\rightarrow0})$ in Eq. (\ref{key-function-2}) is equal to zero,
and then the paramagnetic part of the response kernel (\ref{parakernel}) is reduced as,
\begin{subequations}\label{kernel-6}
\begin{eqnarray}
K^{(\rm p)}_{\hat{x}\hat{x}}(\varsigma,{\bf q}\rightarrow 0,0) &=&  2Z^{(\varsigma)2}_{\rm F}
\frac{4e^2}{\hbar^2}\frac{1}{N}\sum_{\bf k}{\rm sin}^2{k_x}[t_x-2t'{\rm cos}k_y]^{2}\nonumber\\
&\times&\operatorname*{\rm lim}\limits_{{\bf q}\rightarrow 0}
\frac{n_{\rm F}({E^{(\varsigma)}_{\bf k}})
-n_{\rm F}({E^{(\varsigma)}_{\bf k+q}})}{E^{(\varsigma)}_{\bf k}-E^{(\varsigma)}_{\bf k+q}},\\
K^{(\rm p)}_{\hat{y}\hat{y}}(\varsigma,{\bf q}\rightarrow 0,0) &=&  2Z^{(\varsigma)2}_{\rm F}
\frac{4e^2}{\hbar^2}\frac{1}{N}\sum_{\bf k}{\rm sin}^2{k_y}[t_y-2t'{\rm cos}k_x]^{2}\nonumber\\
&\times&\operatorname*{\rm lim}\limits_{{\bf q}\rightarrow 0}
\frac{n_{\rm F}({E^{(\varsigma)}_{\bf k}})
-n_{\rm F}({E^{(\varsigma)}_{\bf k+q}})}{E^{(\varsigma)}_{\bf k}-E^{(\varsigma)}_{\bf k+q}}.
~~~~~~~~~
\end{eqnarray}
\end{subequations}
In this case, the Meissner effect can be discussed respectively in three different temperature
regions: \\
(i) The region at the temperature $T=0$, where a straightforward calculation indicates that
the paramagnetic part of the response kernel in Eq. (\ref{kernel-6}) is equal to zero in the
thermodynamic limit $N\to\infty$, i.e.,
$K^{(\rm p)}_{\hat{x}\hat{x}}(\varsigma,{\bf q}\rightarrow 0,0)|_{T=0}=0$ and
$K^{(\rm p)}_{\hat{y}\hat{y}}(\varsigma,{\bf q}\rightarrow 0,0)|_{T=0}=0$, reflecting a fact
that as in the case of the absence of the electronic nematicity \cite{Liu20}, the zero
temperature electromagnetic response of cuprate superconductors with coexisting electronic
nematicity in the long wavelength limit is also determined by the diamagnetic part of the
response kernel only, i.e., $K_{\hat{x}\hat{x}}(\varsigma,{\bf q},0)$ and
$K_{\hat{y}\hat{y}}(\varsigma,{\bf q},0)$ in Eq. (\ref{kernel-5}) are reduced as,
\begin{subequations}\label{kernel-25}
\begin{eqnarray}
K_{\hat{x}\hat{x}}(\varsigma,{\bf q}\rightarrow 0,0)|_{T=0}&=&
K_{\hat{x}\hat{x}}^{({\rm d})}(\varsigma,{\bf q}\rightarrow 0,0)|_{T=0}\nonumber\\
&=&{1\over\mu_{0}\lambda^{2}_{La}(\varsigma,0)},\\
K_{\hat{y}\hat{y}}(\varsigma,{\bf q\rightarrow 0},0)|_{T=0}&=&
K_{\hat{y}\hat{y}}^{({\rm d})}(\varsigma,{\bf q}\rightarrow 0,0)|_{T=0}\nonumber\\
&=&{1\over\mu_{0}\lambda^{2}_{Lb}(\varsigma,0)},~~~~~~~
\end{eqnarray}
\end{subequations}
respectively.
\begin{figure}
\centering
\includegraphics[scale=0.90]{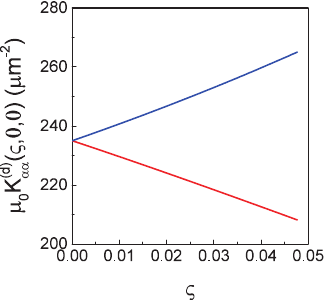}
\caption{(Color online) The diamagnetic part of the response kernel along the $\hat{a}$-axis
(red-line) and $\hat{b}$-axis (blue-line) as a function of the nematic-order state strength at
$\delta=0.15$ with $T=0$. $K_{\alpha\alpha}^{({\rm d})}(0,0)$ is the diamagnetic part of the
response kernel in the case of the absence of the electronic nematicity. \label{London-depth}}
\end{figure}
In Fig. \ref{London-depth}, we plot $K_{\hat{x}\hat{x}}^{({\rm d})}(\varsigma,0,0)$ (red-line)
and $K_{\hat{y}\hat{y}}^{({\rm d})}(\varsigma,0,0)$ (blue-line) as a function of the nematic-order
state strength at doping $\delta=0.15$ with temperature $T=0$. Apparently, the zero-temperature
diamagnetic part of the response kernel $K_{\hat{x}\hat{x}}^{({\rm d})}(\varsigma,0,0)$ along the
$\hat{a}$-axis decreases smoothly with the increase of the strength of the electronic nematicity,
while the zero-temperature diamagnetic part of the response kernel
$K_{\hat{y}\hat{y}}^{({\rm d})}(\varsigma,0,0)$ along the $\hat{b}$-axis rises linearly upon with
the increase of the strength of the electronic nematicity. This anisotropic feature therefore
indicates that the electromagnetic response is inequivalent along the $\hat{a}$- and
$\hat{b}$-axes. \\
\begin{figure*}[t!]
\centering
\includegraphics[scale=0.80]{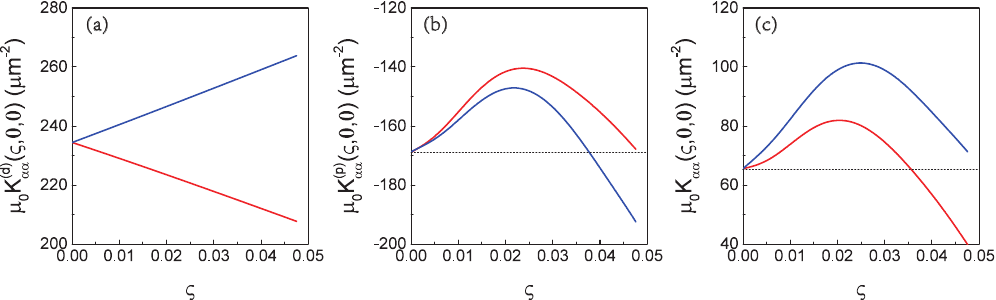}
\caption{(Color online) (a) The diamagnetic part of the response kernel along the $\hat{a}$-axis
(red-line) and $\hat{b}$-axis (blue-line), (b) the paramagnetic part of the response kernel along
the $\hat{a}$-axis (red-line) and $\hat{b}$-axis (blue-line), and (c) the response kernel along
the $\hat{a}$-axis (red-line) and $\hat{b}$-axis (blue-line) as a function of the nematic-order
state strength at $\delta=0.15$ with $T=0.025J$. $K_{\alpha\alpha}^{({\rm d})}(0,0)$,
$K_{\alpha\alpha}^{({\rm p})}(0,0)$, and $K_{\alpha\alpha}(0,0)$ are the diamagnetic and
paramagnetic parts of the response kernel, and the response kernel in the case of the absence of
the electronic nematicity. \label{LP-depth-ab}}
\end{figure*}
(ii) The region at the temperature $0<T<T_{\rm c}$, where the Meissner effect is determined by
both the diamagnetic and paramagnetic parts of the response kernel. In Fig. \ref{LP-depth-ab},
we plot (a) $K_{\hat{x}\hat{x}}^{({\rm d})}(\varsigma,0,0)$ (red-line) and
$K_{\hat{y}\hat{y}}^{({\rm d})}(\varsigma,0,0)$ (blue-line), (b)
$K_{\hat{x}\hat{x}}^{({\rm p})}(\varsigma,0,0)$ (red-line) and
$K_{\hat{y}\hat{y}}^{({\rm p})}(\varsigma,0,0)$ (blue-line), (c)
$K_{\hat{x}\hat{x}}(\varsigma,0,0)$ (red-line) and $K_{\hat{y}\hat{y}}(\varsigma,0,0)$
(blue-line) as a function of the nematic-order state strength at $\delta=0.15$ with
$T=0.025J$, where the typical features can be summarized as: (A) the global feature of the
diamagnetic part of the response kernel along the $\hat{a}$-axis (the $\hat{b}$-axis) at a finite
temperature is the same as that along the $\hat{a}$-axis (the $\hat{b}$-axis) at zero temperature;
(B) although the value of the paramagnetic part of the response kernel along the $\hat{a}$-axis
($\hat{b}$-axis) is negative, it has a dome-like shape nematic-order state strength
dependence; (C) as a result of the sum of the corresponding diamagnetic and paramagnetic parts,
the response kernel along the $\hat{a}$-axis (the $\hat{b}$-axis) exhibits a dome-like shape
nematic-order state strength dependence. In particular, $K_{\hat{x}\hat{x}}(\varsigma,0,0)$ and
$K_{\hat{y}\hat{y}}(\varsigma,0,0)$ are a increasing function of the nematic-order state
strength, the system is thought to be at the lower strength region. The system is at around the
{\it critical strength region}, where $K_{\hat{x}\hat{x}}(\varsigma,0,0)$ and
$K_{\hat{y}\hat{y}}(\varsigma,0,0)$ reach their maximums at around
$\varsigma\approx 0.021$ and $\varsigma\approx 0.025$, respectively. However, with the further
increase in the strength, $K_{\hat{x}\hat{x}}(\varsigma,0,0)$ and
$K_{\hat{y}\hat{y}}(\varsigma,0,0)$ decrease at the higher strength region. Moreover, in the
extremely high strength region, $K_{\hat{x}\hat{x}}(\varsigma,0,0)$ and
$K_{\hat{y}\hat{y}}(\varsigma,0,0)$ are less than those in the case of the absence of the
electronic nematicity \cite{Liu20}.

On the other, in the region at the temperature $0<T<T_{\rm c}$, the magnetic-field
penetration-depths along the $\hat{a}$- and $\hat{b}$-axes are defined as,
\begin{subequations}\label{kernel-28}
\begin{eqnarray}
{1\over\lambda^{2}_{a}(\varsigma,T)}&=& \mu_{0}
K_{\hat{x}\hat{x}}(\varsigma,{\bf q}\rightarrow 0,0),\\
{1\over\lambda^{2}_{b}(\varsigma,T)}&=& \mu_{0}
K_{\hat{y}\hat{y}}(\varsigma,{\bf q}\rightarrow 0,0),
\end{eqnarray}
\end{subequations}
respectively, which can also be used for a direct comparison with the corresponding
experimental results in the clean-limit \cite{Schrieffer64}. This magnetic-field
penetration-depth $\lambda_{a}(\varsigma,T)$ [$\lambda_{b}(\varsigma,T)$] along the
$\hat{a}$-axis [$\hat{b}$-axis] characterizes the length scale along the $\hat{a}$-axis
[$\hat{b}$-axis] over which the supercurrent in cuprate superconductors screens out an
external magnetic-field. The results in Eq. (\ref{kernel-28}) indicate that in a striking
contrast to the case of the nematic-order state strength dependence of the response kernel
shown in Fig. \ref{LP-depth-ab}c, the magnetic-field penetration-depth exhibits a remarkably
reverse dome-like shape of the nematic-order state strength dependence. Moreover, the obtained
results from Eq. (\ref{kernel-28}) also show that at the temperature $T=0.025J$, the
magnetic-field penetration-depths $\lambda_{a}(\varsigma,T)\approx 221.0$ nm along the
$\hat{a}$-axis and $\lambda_{b}(\varsigma,T)\approx 205.4$ nm along the $\hat{b}$-axis at
$\delta=0.15$ for the nematic-order state strength $\varsigma=0.022$, which are consistent
with the experimental results \cite{Quijada99} of $\lambda_{a}=196$ nm and $\lambda_{b}=180$ nm,
respectively, observed on the optimally doped Bi$_{2}$Sr$_{2}$CaCu$_{2}$O$_{8+\delta}$. This
inequivalence between $\lambda_{a}(\varsigma,T)$ along the $\hat{a}$-axis and
$\lambda_{b}(\varsigma,T)$ along the $\hat{b}$-axis therefore further verifies the rotation
symmetry breaking in the electromagnetic response of cuprate superconductors with coexisting
electronic nematicity.

Superconductivity requires that both the electron pair formation and macroscopic phase
coherence happen simultaneously at $T_{\rm c}$, where the phase coherence is controlled by
the superfluid density, associated with the magnetic-field penetration-depth. However, the
inequivalence between $\lambda_{a}(\varsigma,T)$ along the $\hat{a}$-axis and
$\lambda_{b}(\varsigma,T)$ along the $\hat{b}$-axis also induces an inequivalence between
the superfluid densities $\rho^{(a)}_{\rm s}(\varsigma,T)$ along the $\hat{a}$-axis and
$\rho^{(b)}_{\rm s}(\varsigma,T)$ along the $\hat{b}$-axis, where
$\rho^{(a)}_{\rm s}(\varsigma,T)$ and $\rho^{(b)}_{\rm s}(\varsigma,T)$ are identical
respectively to the inverse of the $\lambda_{a}(\varsigma,T)$ square and
$\lambda_{b}(\varsigma,T)$ square in Eq. (\ref{kernel-28}) as,
\begin{subequations}\label{superfluid-density}
\begin{eqnarray}
\rho^{(a)}_{\rm s}(\varsigma,T)&\equiv& {1\over \lambda_{a}^{2}(\varsigma,T)},
\label{superfluid-density-x}\\
\rho^{(b)}_{\rm s}(\varsigma,T)&\equiv& {1\over \lambda_{b}^{2}(\varsigma,T)}.
\label{superfluid-density-y}
\end{eqnarray}
\end{subequations}
The above obtained results in Eqs. (\ref{kernel-28}) and (\ref{superfluid-density}) therefore
show that the superfluid density along the $\hat{a}$-axis ($\hat{b}$-axis) presents a similar
behavior of the response kernel along the $\hat{a}$-axis ($\hat{b}$-axis) shown in
Fig. \ref{LP-depth-ab}c. However, in this paper, the main purpose is to investigate the
evolution of the electromagnetic response with the strength of the electronic nematicity. In
this case, a more appropriate quantities for the depiction of the anomalous form of the
superfluid density as a function of the nematic-order state strength is the average superfluid
density, which is defined as,
\begin{eqnarray}\label{average-density}
\bar{\rho}_{\rm s}(\varsigma,T)=\sqrt{\rho^{(a)}_{{\rm s}}(\varsigma,T)
\rho^{(b)}_{{\rm s}}(\varsigma,T)}.
\end{eqnarray}
\begin{figure}
\centering
\includegraphics[scale=0.75]{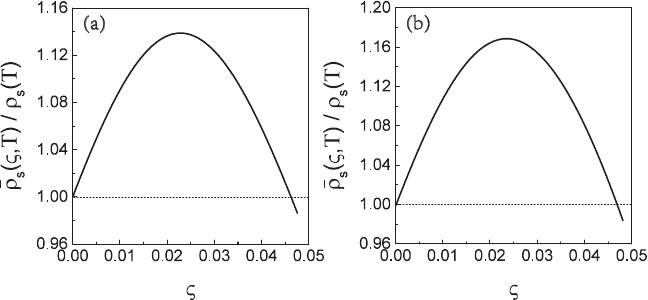}
\caption{The average superfluid density as a function of the nematic-order state
strength at (a) $\delta=0.15$ and (b) $\delta=0.12$ with $T=0.025J$. $\rho_{\rm s}(T)$ is the
superfluid density in the case of the absence of the electronic nematicity.
\label{density-strenth}}
\end{figure}
To show the exotic behavior of the nematic-order state strength dependence of the average
superfluid density $\bar{\rho}_{\rm s}(\varsigma,T)$ more clearly, we plot
$\bar{\rho}_{\rm s}(\varsigma,T)$ as a function of the nematic-order state strength at (a)
$\delta=0.15$ and (b) $\delta=0.12$ with $T=0.025J$ in Fig. \ref{density-strenth}. One can
immediately see from the results in Fig. \ref{density-strenth} that
$\bar{\rho}_{\rm s}(\varsigma,T)$ presents a dome-like shape nematic-order state strength
dependence, where a distinct peak appears at around the {\it critical strength} of the
electronic nematicity $\varsigma_{\rm critical}\approx 0.023$, and then when the strength of
the electronic nematicity is tuned away from the critical strength, this pronounced peak is
suppressed at the lower strength as well as at the higher strength sides. More importantly,
the strength range together with the {\it critical strength} of $\bar{\rho}_{\rm s}(\varsigma,T)$
at the underdoping $\delta=0.12$ are the exact same with those at the optimal doping
$\delta=0.15$, indicating that the dome-like shape of the nematic-order state strength
dependence of $\bar{\rho}_{\rm s}(\varsigma,T)$ occurs at a any given doping of the SC dome. This
in turn leads to the enhancement of superconductivity \cite{Cao22,Cao21}, and gives rise to
the dome-like shape of the nematic-order state strength dependence of $T_{\rm c}$. However,
in the extremely high strength region $\varsigma> 0.045$, $\bar{\rho}_{\rm s}(\varsigma,T)$
is less than that in the case of the absence of the electronic nematicity \cite{Liu20}, which
leads to a reduction of $T_{\rm c}$.\\
(iii) The region at the temperature $T=T_{\rm c}$, where the SC gap
$\bar{\Delta}^{(\varsigma)}({\bf k})|_{T=T_{\rm c}}=0$. Following our previous discussions in
the case of the absence of the electronic nematicity \cite{Liu20}, the paramagnetic part of
the response kernel in Eq. (\ref{kernel-6}) can be reduced as,
\begin{subequations}
\begin{eqnarray}
K_{\hat{x}\hat{x}}^{({\rm p})}(\varsigma,{\bf q}\rightarrow 0,0)|_{T=T_{\rm c}}&=&
-K_{\hat{x}\hat{x}}^{({\rm d})}(\varsigma,{\bf q}\rightarrow 0,0)|_{T=T_{\rm c}}\nonumber\\
&=&-{1\over\mu_{0}\lambda^{2}_{La}(\varsigma,T)}|_{T=T_{\rm c}},\\
K_{\hat{y}\hat{y}}^{({\rm p})}(\varsigma,{\bf q}\rightarrow 0,0)|_{T=T_{\rm c}}&=&
-K_{\hat{y}\hat{y}}^{({\rm d})}(\varsigma,{\bf q}\rightarrow 0,0)|_{T=T_{\rm c}}\nonumber\\
&=&-{1\over\mu_{0}\lambda^{2}_{Lb}(\varsigma,T)}|_{T=T_{\rm c}},~~~~~~~
\end{eqnarray}
\end{subequations}
which exactly cancel the corresponding diamagnetic part of the response kernel in
Eqs. (\ref{diakernel-x}) and (\ref{diakernel-y}), respectively, reflecting a basic fact that
the Meissner effect in cuprate superconductors with coexisting electronic nematicity occurs
below $T_{\rm c}$ only.

In summary, we have found the following results within the kinetic-energy-driven
superconductivity: (i) the Meissner effect in cuprate superconductors with coexisting electronic
nematicity is obtained for all temperature $T\leq T_{c}$; (ii) the electromagnetic response is
inequivalent along with the $\hat{a}$- and $\hat{b}$-axes; (iii) the response kernels along the
$\hat{a}$- and $\hat{b}$-axes are not manifestly gauge invariant within the bare current vertex
in Eq. (\ref{barevertex}), however, the gauge invariance can be kept within the dressed current
vertex \cite{Feng15}.

\subsection{Local magnetic-field profile with broken rotation symmetry}
\label{local-profile}

We now turn to derive the local magnetic-field profile based on the standard specular reflection
model with a two-dimensional geometry \cite{Abrikosov88,Tinkham96}. The local magnetic-field
profile can be measured experimentally, e.g., by using the muon-spin rotation technique
\cite{Jackson00,Khasanov04,Suter04}, reflecting the electromagnetic response and yielding the
crucial information of the magnetic-field screening inside the sample. In cuprate
superconductors, the experimental observations indicates an exponential character of the
magnetic-field screening \cite{Jackson00,Khasanov04,Suter04}, in support of a local nature of
the electrodynamics \cite{Schrieffer64}. However, the rotation symmetry-breaking of the
response kernel (\ref{kernel-5}) is inequivalent along the $\hat{a}$- and $\hat{b}$-axes. In
this case, if the external magnetic-field is perpendicular to the ab plane, we can choose
$A_{y}(x)$ along the $\hat{b}$-axis, or $A_{x}(y)$ along the $\hat{a}$-axis. From the
following Maxwell equation,
\begin{eqnarray}
{\rm rot}\,{\bf B}={\rm rot}\,{\rm rot}\,{\bf A}={\rm grad}\,{\rm div}\,{\bf A}
-\nabla^{2}\,{\bf A}=\mu_{0}{\bf J},
\end{eqnarray}
it can be found that the extension of the vector potential in an even manner through the
boundary implies a kink in the $A_{y}(x)$ $[A_{x}(y)]$ curve. In other words, if the external
magnetic field ${\bf B}$ is given at the system surface, i.e.,
$({\rm d}A_{y}(x)/{\rm d}x)|_{x=+0}=B$, while $({\rm d}A_{y}(x)/{\rm d}x)|_{x=-0} =-B$, or
$({\rm d}A_{x}(y)/{\rm d}y)|_{y=+0}=B$, while $({\rm d}A_{x}(y)/{\rm d}y)|_{y=-0} =-B$, which
\cite{Abrikosov88} indicates that the second derivative $({\rm d}^{2}A_{y}(x)/ {\rm d}^{2}x)$
acquires a correction $2B\delta(x)$, or $({\rm d}^{2}A_{x}(y)/{\rm d}^{2}y)$ acquires a
correction $2B\delta(y)$,
\begin{subequations}\label{correction}
\begin{eqnarray}
{{\rm d}^{2}A_{y}(x)\over {\rm d}^{2}x}=2B\delta(x)-\mu_{0}J^{(\varsigma)}_{y}, \\
{{\rm d}^{2}A_{x}(y)\over {\rm d}^{2}y}=2B\delta(y)-\mu_{0}J^{(\varsigma)}_{x},
\end{eqnarray}
\end{subequations}
where the transverse gauge ${\rm div}\,{\bf A}=0$ has been adopted. In the momentum space, the
above these equations can be expressed as,
\begin{subequations}\label{FT-correction}
\begin{eqnarray}
q_{x}^{2}A_{y}({\bf q})=\mu_{0}J^{(\varsigma)}_{y}({\bf q})-2B, \\
q_{y}^{2}A_{x}({\bf q})=\mu_{0}J^{(\varsigma)}_{x}({\bf q})-2B.
\end{eqnarray}
\end{subequations}
Substituting this Fourier transform form (\ref{FT-correction}) into Eq. (\ref{linres}), and
performing a solution for the vector potential, the relations between the vector potential and
the response kernels can be obtained as,
\begin{subequations}\label{aspec}
\begin{eqnarray}
A_{y}({\bf q})=-2B{\delta(q_{y})\delta(q_{z})\over\mu_{0}
K_{\hat{y}\hat{y}}(\varsigma,{\bf q})+q_{x}^{2}}, \\
A_{x}({\bf q})=-2B{\delta(q_{x})\delta(q_{z})\over\mu_{0}
K_{\hat{x}\hat{x}}(\varsigma,{\bf q})+q_{y}^{2}}.
\end{eqnarray}
\end{subequations}
Since the vector potential has only the $\hat{y}$ [$\hat{x}$] component, the non-zero component
of the local magnetic-field ${\bf h}=\rm{rot}\,{\bf A}$ is that along the $z$ axis as
$h^{(\varsigma)}_{z x}({\bf q})=iq_{x}A_{y}({\bf q})$
[$h^{(\varsigma)}_{z y}({\bf q}) =iq_{y}A_{x}({\bf q})$].

With the help of the above relations in Eq. (\ref{aspec}) and the response kernels in
Eq. (\ref{kernel-28}), the local magnetic-field profiles along the $\hat{a}$- and $\hat{b}$-axes
in the long wavelength limit can be derived straightforwardly as,
\begin{subequations}\label{profile}
\begin{eqnarray}
h^{(\varsigma)}_{z}(x)&=&{B\over\pi}\int\limits_{-\infty}^\infty {\rm{d}}q_{x}\,{q_{x}
\sin(q_{x}x)\over\mu_{0}K_{\hat{y}\hat{y}}(\varsigma,q_{x}\rightarrow 0,0,0)+q_{x}^{2}}\nonumber\\
&=&2Be^{-{x\over\lambda_{b}(\varsigma,T)}},~~~~\\
h^{(\varsigma)}_{z}(y)&=&{B\over\pi}\int\limits_{-\infty}^\infty {\rm{d}}q_{y}\,{q_{y}
\sin(q_{y}y)\over\mu_{0}K_{\hat{x}\hat{x}}(\varsigma,q_{y}\rightarrow 0,0,0)+q_{y}^{2}}\nonumber\\
&=&2Be^{-{y\over\lambda_{a}(\varsigma,T)}},~~~~
\end{eqnarray}
\end{subequations}
respectively. In a striking analogy to the case of the absence of the electronic nematicity
\cite{Liu20}, the distance dependence of $h^{(\varsigma)}_{z}(x)$ [$h^{(\varsigma)}_{z}(y)$)]
follows an exponential law as was expected for the local electrodynamic response. However, the
magnitude of $h^{(\varsigma)}_{z}(x)$ along the $\hat{a}$-axis at a given distance is unequal
to the corresponding one of $h^{(\varsigma)}_{z}(y)$ along the $\hat{b}$-axis, with the
difference of the magnitudes between $h^{(\varsigma)}_{z}(x)$ and $h^{(\varsigma)}_{z}(y)$ that
is increased with the increase of distance, in qualitative agreement with the experimental
results \cite{Kiefl10}. This anisotropic feature therefore reflects an experimental fact that
the electromagnetic response is inequivalent along the $\hat{a}$- and $\hat{b}$-axes
\cite{Kiefl10}.

The enhancement of the superfluid density by the electronic nematicity can be attributed to the
enhancement of the SC condensation energy, i.e., the energy of the system in the SC-state with
coexisting nematic order is lower than the energy in the SC-state with the absence of the nematic
order. In other words, the SC-state with coexisting nematic order is more stable than the SC-state
with the absence of the nematic order. The internal energy $U^{\rm (s)}_{\varsigma}(T)$ of the
system can be expressed as,
\begin{eqnarray}\label{internal-energy-1}
U^{\rm (s)}_{\varsigma}(T)=2\int\limits_{-\infty}^\infty {{\rm{d}}\omega\over 2\pi}
\rho^{\rm (s)}(\omega,T,\varsigma)\omega n_{\rm F}(\omega),~~~~
\end{eqnarray}
with the fermion distribution function $n_{\rm F}(\omega)$, and the electron density of states
$\rho^{\rm (s)}(\omega,T,\varsigma)$,
\begin{eqnarray}\label{electron-density}
\rho^{\rm (s)}(\omega,T,\varsigma)={1\over N}\sum_{{\bf k}}A_{\varsigma}({\bf k},\omega,T),~~~
\end{eqnarray}
where the electron spectral function
$A_{\varsigma}({\bf k},\omega,T)=-2{\rm Im}G^{\rm (RMF)}_{\varsigma}({\bf k},\omega)$ is
obtained directly from the electron diagonal propagator
$G^{\rm (RMF)}_{\varsigma}({\bf k},\omega)$ in Eq. (\ref{NPEGF}) as,
\begin{eqnarray}\label{spectral-function}
A_{\varsigma}({\bf k},\omega,T)&=&\pi Z^{(\varsigma)}_{\rm F}\left [ \left ( 1
+{\bar{\varepsilon}^{(\varsigma)}_{\bf k}\over E^{(\varsigma)}_{\bf k}}\right )
\delta(\omega-E^{(\varsigma)}_{\bf k})\right .\nonumber\\
&+&\left. \left ( 1-{\bar{\varepsilon}^{(\varsigma)}_{\bf k}\over
E^{(\varsigma)}_{\bf k}}\right )\delta(\omega+E^{(\varsigma)}_{\bf k}) \right ]. ~~~
\end{eqnarray}
Substituting $A_{\varsigma}({\bf k},\omega,T)$ in Eq. (\ref{spectral-function}) into
Eqs. (\ref{electron-density}) and (\ref{internal-energy-1}), $U^{\rm (s)}_{\varsigma}(T)$ can
be evaluated as,
\begin{eqnarray}\label{internal-energy-sc}
U^{(\rm s)}_{\varsigma}(T)=-{Z^ {(\varsigma)}_{{\rm F}}\over N}\sum_{\bf k}
[E^{(\varsigma)}_{\bf k}{\rm tanh}({1\over 2}\beta E^{(\varsigma)}_{\bf k}]
+{Z^ {(\varsigma)}_{{\rm F}}\over N}\sum_{\bf k}\bar{\varepsilon}^{(\varsigma)}_{\bf k}.~~~~~
\end{eqnarray}
In the normal-state, the SC gap $\bar{\Delta}^{(\varsigma)}({\bf k})=0$, this internal energy
is reduced as,
\begin{eqnarray}\label{internal-energy-nor}
U^{(\rm n)}_{\varsigma}(T)=-{Z^ {(\varsigma)}_{{\rm F}}\over N}\sum_{\bf k}
[\bar{\varepsilon}^{(\varsigma)}_{\bf k}
{\rm tanh}({1\over 2}\beta\bar{\varepsilon}^{(\varsigma)}_{\bf k})]
+{Z^ {(\varsigma)}_{{\rm F}}\over N}\sum_{\bf k}\bar{\varepsilon}^{(\varsigma)}_{\bf k}.~~~~~
\end{eqnarray}
At zero temperature, the SC condensation energy $E^{({\varsigma})}_{\rm cond}(T)$ can be
obtained as,
\begin{eqnarray}\label{condensation-energy}
E^{(\varsigma)}_{\rm cond}&=&U^{(\rm n)}_{{\varsigma}}(T)-U^{(\rm s)}_{{\varsigma}}(T)|_{T=0}.
~~~~~
\end{eqnarray}

\begin{figure}
\centering
\includegraphics[scale=0.90]{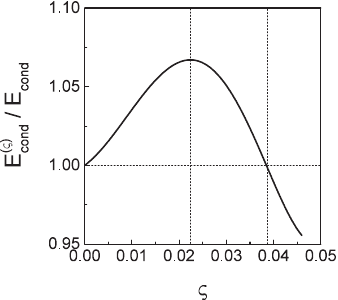}
\caption{The superconducting condensation energy as a function of the nematic-order state
strength at $\delta=0.15$. $E_{\rm cond}$ is the superconducting condensation energy in the
case of the absence of the electronic nematicity. \label{energy-nematicity}}
\end{figure}

\begin{figure*}[t!]
\centering
\includegraphics[scale=0.80]{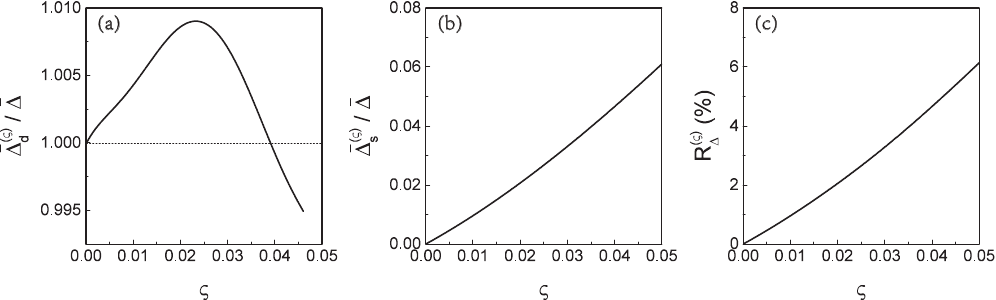}
\caption{(a) The d-wave component of the superconducting gap parameter, (b) the s-wave component
of the superconducting gap parameter, and (c) the ratio of the s-wave component to d-wave
component as a function of the nematic-order state strength at $\delta=0.15$ with $T=0$.
$\bar{\Delta}$ is the superconducting gap parameter in the case of the absence of the electronic
nematicity. \label{gap-parameter}}
\end{figure*}

In Fig. \ref{energy-nematicity}, we plot $E^{(\varsigma)}_{\rm cond}$ as a function of the
nematic-order state strength at $\delta=0.15$, where there is the surprising similarity
between the $E^{(\varsigma)}_{\rm cond}$ and $\bar{\rho}_{\rm s}(\varsigma)$ with the
following characteristic features: (i) $E^{(\varsigma)}_{\rm cond}$ is enhanced by the
electronic nematicity in the whole strength range of the electronic nematicity except for in
the the extremely strong strength region $\varsigma> 0.04$, where $E^{(\varsigma)}_{\rm cond}$
is reduced. This enhancement of $E^{(\varsigma)}_{\rm cond}$ therefore induces the enhancement
of $\bar{\rho}_{\rm s}(\varsigma,T)$. The result in Fig. \ref{energy-nematicity} also reflects
a fact that the strong electron correlation induces the system to find new way to lower its
ground-state energy by the spontaneous breaking of the native rotation symmetry of the square
lattice underlying the copper-oxide plane \cite{Kivelson19,Vojta09,Fradkin10,Fernandes19}. In
other words, as the appearance of superconductivity, the emergence of the electronic nematicity
together with the associated fluctuation phenomena in the whole strength range of the
electronic nematicity except for in the extremely strong strength region are a natural
consequence of the strong electron correlation effect; (ii) However, there is a substantial
difference, namely, $E^{(\varsigma)}_{\rm cond}$ less than that in the case of the absence of
the electronic nematicity occurs at the extremely strong strength region $\varsigma> 0.04$,
rather than at the extremely high strength region $\varsigma> 0.045$, where
$\bar{\rho}_{\rm s}(\varsigma,T)$ is less than that in the case of the absence of the
electronic nematicity, although the crossover strength $\varsigma_{\rm essr}=0.04$ in
$E^{(\varsigma)}_{\rm cond}$ is not far from the crossover strength
$\varsigma_{\rm ehsr}=0.045$ in $\bar{\rho}_{\rm s}(\varsigma,T)$. However, the actual weak
strength region of the order of the magnitude of the strength $\varsigma$ with the high impacts
on various properties \cite{Nakata18,Sato17,Daou10} is exactly same in
$E^{(\varsigma)}_{\rm cond}$ and $\bar{\rho}_{\rm s}(\varsigma,T)$. Moreover, the optimal
strength $\varsigma_{\rm optimal}\approx 0.022$ for the maximal $E^{(\varsigma)}_{\rm cond}$
is quite close to the critical strength $\varsigma_{\rm critical}\approx 0.023$ for the highest
$\bar{\rho}_{\rm s}(\varsigma,T)$.

We now turn to show (i) why $E^{({\varsigma})}_{\rm cond}$ has a dome-like shape of the
nematic-order state strength dependence? and (ii) why $E^{({\varsigma})}_{\rm cond}$ is reduced
by the electronic nematicity in the extremely strong strength region? The expression form of
the SC condensation energy in Eq. (\ref{condensation-energy}) also indicates that
$E^{({\varsigma})}_{\rm cond}$ is proportional to the SC gap, i.e.,
$E^{({\varsigma})}_{\rm cond}(T)\propto \bar{\Delta}^{(\varsigma)}$. The SC gap measures the
strength of the binding of electrons into electron pairs
\cite{Schrieffer64,Bonn96,Sonier00,Basov05,Sonier16}, while the superfluid density is a measure
of the phase stiffness \cite{Schrieffer64,Bonn96,Sonier00,Basov05,Sonier16}, therefore the
SC gap and superfluid density separately describe the different aspects of the same SC
quasiparticles. In the case of the absence of the electronic nematicity \cite{Feng15a}, the pure
d-wave electron pairs with the electron pair strength $\bar{\Delta}\gamma^{\rm (d)}_{\bf k}$
condensation reveals the SC-state with the pure d-wave symmetry. However, the present result of
Eq. (\ref{SCGF-DS}) in the SC-state with coexisting nematic order shows that in addition to the
pure d-wave component of the SC gap $\bar{\Delta}^{(\varsigma)}_{\rm d}\gamma^{\rm (d)}_{\bf k}$,
the pure s-wave component of the SC gap
$\bar{\Delta}^{(\varsigma)}_{\rm s}\gamma^{\rm (s)}_{\bf k}$ is induced by the electronically
nematic order, therefore there is a coexistence and competition between the d-wave component of
the SC gap parameter $\bar{\Delta}^{(\varsigma)}_{\rm d}$ and the s-wave component of the SC gap
parameter $\bar{\Delta}^{(\varsigma)}_{\rm s}$. This coexistence and competition is closely
related to the strength of the electronic nematicity, and therefore plays a crucial role in the
exotic features of the nematic-order state strength dependence of the electromagnetic response.
To show this point more clearly, we plot (a) the d-wave component of the SC gap parameter
$\bar{\Delta}^{(\varsigma)}_{\rm d}$, (b) the s-wave component of the SC gap parameter
$\bar{\Delta}^{(\varsigma)}_{\rm s}$, and (c) the ratio of the s-wave component to d-wave
component
$R^{(\varsigma)}_{\Delta}=\bar{\Delta}^{(\varsigma)}_{\rm s}/\bar{\Delta}^{(\varsigma)}_{\rm d}$
as a function of the nematic-order state strength at $\delta=0.15$ with $T=0$ in
Fig. \ref{gap-parameter}, where the key features can be summarized as: (i)
$\bar{\Delta}^{(\varsigma)}_{\rm d}$ exhibits a dome-like shape nematic-order state strength
dependence (see Fig. \ref{gap-parameter}a), while the result shows a almost linear
characteristics of $\bar{\Delta}^{(\varsigma)}_{\rm s}$ (see Fig. \ref{gap-parameter}b). In
other words, $\bar{\Delta}^{(\varsigma)}_{\rm s}$ increases monotonically with the increase of
the nematic-order state strength in the whole strength range of the electronic nematicity,
while $\bar{\Delta}^{(\varsigma)}_{\rm d}$ is enhanced by the electronic nematicity from the
weak to strong strength regions of the electronic nematicity except for in the the extremely
strong strength region, where $\bar{\Delta}^{(\varsigma)}_{\rm d}$ is reduced. In particular,
the nematic-order state strength range of the SC dome together with the {\it optimal} strength
in $\bar{\Delta}^{(\varsigma)}_{\rm d}$ are almost the same with those in
$\bar{\rho}_{\rm s}(\varsigma,T)$, which is an evidence that the nematic-order state strength
dependence of $\bar{\rho}_{\rm s}(\varsigma,T)$ is mainly determined by the nematic-order
state strength dependence of $\bar{\Delta}^{(\varsigma)}_{\rm d}$; (ii) In the lower ratio
region ($0<R^{(\varsigma)}_{\Delta}< 2.29\%$, see Fig. \ref{gap-parameter}c), which is
corresponding to the weak strength region of the electronic nematicity ($\varsigma< 0.022$),
the electronic nematicity induces an increase of both $\bar{\Delta}^{(\varsigma)}_{\rm d}$
and $\bar{\Delta}^{(\varsigma)}_{\rm s}$. In particular, although the increase rate for
$\bar{\Delta}^{(\varsigma)}_{\rm d}$ is slower than that in $\bar{\Delta}^{(\varsigma)}_{\rm s}$,
the maximal SC gap parameter $\bar{\Delta}^{(\varsigma)}$ in Eq. (\ref{SCGF-DS}) is increased,
which leads to that $\bar{\rho}_{\rm s}(\varsigma,T)$ increase with the increase of the
nematic-order state strength in the lower ratio region. However, in the higher ratio region
($4.66\%>R^{(\varsigma)}_{\Delta}> 2.29\%$), which is corresponding to the strong strength
region of the electronic nematicity ($0.022<\varsigma < 0.04$), the electronic nematicity tends
to support the high speed increase of $\bar{\Delta}^{(\varsigma)}_{\rm s}$, concomitantly,
$\bar{\Delta}^{(\varsigma)}_{\rm d}$ is decreased. In this case, the maximal SC gap parameter
$\bar{\Delta}^{(\varsigma)}$ in Eq. (\ref{SCGF-DS}) is decreased, which leads to that
$\bar{\rho}_{\rm s}(\varsigma,T)$ decrease with the increase of the nematic-order state strength
in the higher ratio region. The optimal ratio region ($R^{(\varsigma)}_{\Delta}\sim 2.29\%$),
corresponding to the optimal strength of the electronic nematicity ($\varsigma = 0.022$), is a
balance region, where both $\bar{\Delta}^{(\varsigma)}_{\rm d}$ and
$\bar{\Delta}^{(\varsigma)}_{\rm s}$ and the nematic-order state strength are optimally matched,
leading to that the highest $\bar{\rho}_{\rm s}(\varsigma,T)$ appears at around the optimal
ratio region. This is why the highest $\bar{\rho}_{\rm s}(\varsigma,T)$ occurs at around the
optimal ratio region, and then decreases in both the lower and higher ratio regions. However,
in the extremely higher ratio region ($R^{(\varsigma)}_{\Delta}> 4.66\%$), which is
corresponding to the extremely strong strength region of the electronic nematicity, the
increased part in $\bar{\Delta}^{(\varsigma)}_{\rm s}$ can not compensate for the lost part in
$\bar{\Delta}^{(\varsigma)}_{\rm d}$, and then the maximal SC gap parameter
$\bar{\Delta}^{(\varsigma)}$ in Eq. (\ref{SCGF-DS}) is less than that in the case of the
absence of the electronic nematicity, which leads to that $\bar{\rho}_{\rm s}(\varsigma,T)$ is
less than that in the case of the absence of the electronic nematicity. This is why
$\bar{\rho}_{\rm s}(\varsigma,T)$ in the extremely strong strength region is lower than that in
the case of the absence of the electronic nematicity.

\section{Summary and discussion}\label{conclusions}

Within the framework of the kinetic-energy-driven superconductivity, we have investigated the
nematic-order state strength dependence of the electromagnetic response in cuprate
superconductors in terms of the linear response approach, where the rotation symmetry-breaking
of the response kernel is evaluated and employed to calculate the magnetic-field
penetration-depth, the superfluid density, and the local magnetic-field profile, for a purely
transverse vector potential. Our results indicate that the electromagnetic response of cuprate
superconductors with coexisting electronic nematicity is inequivalent along the $\hat{a}$- and
$\hat{b}$-axes. In particular, the calculated local-magnetic-field profiles along the $\hat{a}$-
and $\hat{b}$-axes as a function of distance and the magnetic-field penetration-depths along
the $\hat{a}$- and $\hat{b}$-axes for the optimal strength of the electronic nematicity are
qualitatively consistent with the corresponding experimental results \cite{Kiefl10,Quijada99}.
The obtained results also show that in addition to the pure d-wave component of the SC gap, the
pure s-wave component of the SC gap is generated by the electronically nematic order, therefore
there is a coexistence and competition of the pure d-wave component and the pure s-wave component.
However, this coexistence and competition leads to the average superfluid density that first
increases with the strength of the electronic nematicity in the lower strength region, then
reaches a maximum value at around the critical strength of the electronic nematicity, but is
suppressed with further increase of the strength in the higher strength region of the electronic
nematicity, which in turn induces the enhancement of superconductivity, and gives rise to the
dome-like shape of the nematic-order state strength dependence of the superfluid density.

Finally, it should be emphasized that besides the emergence of the electronic nematicity in
cuprate superconductors \cite{Fradkin15,Kivelson19,Vojta09,Fradkin10,Fernandes19}, the
electronically nematic order has been detected from other families of the unconventional
superconductors, including the iron-based superconductors
\cite{Chuang10,Gallais13,Massat16,Baek20}, the strontium ruthenate superconductors \cite{Borzi07},
as well as the nickel-based superconductors \cite{Eckberg20}, and then a characteristic feature
in the complicated phase diagrams of these unconventional superconductors is the interplay
between the electronic nematicity and superconductivity. In this case, the theoretical framework
developed in this paper for the understanding of the nature of the electromagnetic response of
cuprate superconductors with coexisting electronic nematicity can be also employed to study the
electromagnetic response of these unconventional superconductors with coexisting electronic
nematicity \cite{Chuang10,Gallais13,Massat16,Baek20,Borzi07,Eckberg20}. In particular, in the
iron-based superconductors \cite{Chuang10,Gallais13,Massat16,Baek20}, the SC gap with a simple
$s_{\pm}$ symmetry
$|\Delta({\bf k})|=|\Delta_{0}\gamma^{\rm (s)}_{\bf k}|=\Delta_{0}|\cos k_{x}+\cos k_{y}|/2$
in the case of the absence of the electronic nematicity is modified as,
\begin{eqnarray}
|\Delta^{(\varsigma)}_{0}({\bf k})|=|\Delta^{(\varsigma)}_{\rm 0s}\gamma^{\rm (s)}_{\bf k}
+\Delta^{(\varsigma)}_{\rm 0d}\gamma^{\rm (d)}_{\bf k}|, \label{iron-based-gap}
\end{eqnarray}
in the case of the presence of the electronic nematicity, where
$\Delta^{(\varsigma)}_{\rm 0s}=(\Delta^{(\varsigma)}_{0\hat{x}}+\Delta^{(\varsigma)}_{0\hat{y}})/2$
and
$\Delta^{(\varsigma)}_{\rm 0d}=(\Delta^{(\varsigma)}_{0\hat{x}}-\Delta^{(\varsigma)}_{0\hat{y}})/2$
are the s-wave and d-wave components of the SC gap parameter, respectively. It thus shows that this
modification in Eq. (\ref{iron-based-gap}) arising from the emergence of the electronic nematicity
in the iron-based superconductors \cite{Chuang10,Gallais13,Massat16,Baek20} induces a deviation from
the pure $s_{\pm}$ pairing symmetry.


\section*{Acknowledgements}

The authors would like to thank Dr. Yiqun Liu and Dr. Minghuan Zeng for the helpful discussions.

\section*{Disclosure statement}

No potential conflict of interest was reported by the authors.

\section*{Funding}

ZC, XM, and SF are supported by the National Key Research and Development Program of China under
Grant No. 2021YFA1401803, and the National Natural Science Foundation of China (NSFC) under Grant
Nos. 11974051 and 12274036. HG is supported by NSFC under Grant Nos. 11774019 and
12074022, and the Fundamental Research Funds for the Central Universities and HPC resources at
Beihang University.

\begin{appendix}

\section{Electron propagator} \label{Green-function}

This Appendix presents the derivation of the vertex corrected electron propagator
$\mathbb{G}_\varsigma({\bf k},\omega)$ in Eq. (\ref{NPEGF}) of the main text. In the fermion-spin
representation (\ref{CSS}), the original $t$-$J$ model in Eq. (\ref{tJ-model}) at zero magnetic
field can be rewritten as,
\begin{eqnarray}\label{CSS-tJ-model}
H&=&\sum_{l\hat{\eta}}t_{\hat{\eta}}(h^{\dagger}_{l+\hat{\eta}\uparrow}h_{l\uparrow}S^{+}_{l}
S^{-}_{l+\hat{\eta}}+h^{\dagger}_{l+\hat{\eta}\downarrow}h_{l\downarrow}
S^{-}_{l}S^{+}_{l+\hat{\eta}})\nonumber\\
&-& \sum_{l\hat{\eta}'}t'_{\hat{\eta}'}(h^{\dagger}_{l+\hat{\eta}'\uparrow}h_{l\uparrow}S^{+}_{l}
S^{-}_{l+\hat{\eta}'}+h^{\dagger}_{l+\hat{\eta}'\downarrow}h_{l\downarrow}S^{-}_{l}
S^{+}_{l+\hat{\eta}'})\nonumber\\
&-& \mu_{\rm h}\sum_{l\sigma}h^{\dagger}_{l\sigma}h_{l\sigma}+\sum_{l\hat{\eta}}
J^{(\hat{\eta})}_{\rm eff}{\bf S}_{l}\cdot {\bf S}_{l+\hat{\eta}},~~~~~~
\end{eqnarray}
where $\mu_{\rm h}$ is the charge-carrier chemical potential, $S^{-}_{l}$ and $S^{+}_{l}$ are the
spin-lowering and spin-raising operators for the spin $S=1/2$, respectively,
$J_{\rm eff}^{(\hat{\eta})}=(1-\delta)^{2}J_{\hat{\eta}}$ is the effective exchange coupling, and
$\delta=\langle h^{\dagger}_{l\sigma}h_{l\sigma}\rangle$ is the doping concentration.

Within the framework of the kinetic-energy driven superconductivity \cite{Feng15,Feng0306,Feng12},
it has been shown that the interaction between the charge carriers directly from the
kinetic energy of the $t$-$J$ model (\ref{CSS-tJ-model}) by the exchange of a strongly dispersive
spin excitation generates the charge-carrier pairing state with coexisting nematic order
\cite{Cao22,Cao21}, where the charge-carrier diagonal and off-diagonal propagators satisfy the
following self-consistent equations as,
\begin{subequations}\label{CCSCES}
\begin{eqnarray}
g_{\varsigma}({\bf k},\omega)&=&g^{(0)}_{\varsigma}({\bf k},\omega)
+g^{(0)}_{\varsigma}({\bf k},\omega)
[\Sigma^{(\rm h)}_{{\rm ph}}(\varsigma,{\bf k},\omega)g_{\varsigma}({\bf k},\omega)\nonumber\\
&-&\Sigma^{(\rm h)}_{{\rm pp}}(\varsigma,{\bf k},\omega)
\Gamma^{\dagger}_{\varsigma}({\bf k},\omega)], ~~~~~~~\\
\Gamma^{\dagger}_{\varsigma}({\bf k},\omega)&=& g^{(0)}_{\varsigma}({\bf k},-\omega)
[\Sigma^{(\rm h)}_{{\rm ph}}(\varsigma,{\bf k},-\omega)
\Gamma^{\dagger}_{\varsigma}({\bf k},\omega)\nonumber\\
&+&\Sigma^{(\rm h)}_{{\rm pp}}(\varsigma,{\bf k},\omega)g_{\varsigma}({\bf k},\omega)],~~~~
\end{eqnarray}
\end{subequations}
where $g^{(0)}_{\varsigma}({\bf k},\omega)$ is the mean-field (MF) charge-carrier diagonal
propagator, and has been given explicitly in Ref. \onlinecite{Cao22}, while
$\Sigma^{(\rm h)}_{\rm ph}(\varsigma,{\bf k},\omega)$ and
$\Sigma^{(\rm h)}_{\rm pp}(\varsigma,{\bf k},\omega)$ are the charge-carrier self-energies in the
particle-hole and particle-particle channels, respectively, and have been obtained in terms of
the spin bubble as \cite{Cao22,Cao21},
\begin{subequations}\label{CCSE}
\begin{eqnarray}
\Sigma^{({\rm h})}_{{\rm ph}}(\varsigma,{\bf k},i\omega_{n})&=&{1\over N^{2}}
\sum_{{\bf p},{\bf p}'}[\Lambda^{(\varsigma)}_{{\bf p}+{\bf p}'+{\bf k}}]^{2}\nonumber\\
&\times&{1\over\beta}\sum_{ip_{m}}
g_{\varsigma}({{\bf p}+{\bf k}},ip_{m}+i\omega_{n})\Pi_{\varsigma}({\bf p},{\bf p}',ip_{m}),\nonumber\\
~~\\
\Sigma^{({\rm h})}_{{\rm pp}}(\varsigma,{\bf k},i\omega_{n})&=&{1\over N^{2}}
\sum_{{\bf p},{\bf p}'}[\Lambda^{(\varsigma)}_{{\bf p}+{\bf p}'+{\bf k}}]^{2}\nonumber\\
&\times&{1\over\beta}\sum_{ip_{m}}
\Gamma^{\dagger}_{\varsigma}({\bf p}+{\bf k},ip_{m}+i\omega_{n})
\Pi_{\varsigma}({\bf p},{\bf p}',ip_{m}),\nonumber\\
\end{eqnarray}
\end{subequations}
with the fermionic and bosonic Matsubara frequencies $\omega_{n}$ and $p_{m}$, respectively,
the bare vertex function $\Lambda^{(\varsigma)}_{\bf k}=4t[(1-\varsigma)\gamma_{{\bf k}_{x}}
+(1+\varsigma)\gamma_{{\bf k}_{y}}]-4t'\gamma_{\bf{k}}'$, and the spin bubble,
\begin{eqnarray}
\Pi_{\varsigma}({\bf p},{\bf p}',ip_{m})&=&{1\over\beta}\sum_{ip_{m}'}
D^{(0)}_{\varsigma}({\bf p}',ip_{m}')\nonumber\\
&\times&D^{(0)}_{\varsigma}({\bf p'+p},ip_{m}'+ip_{m}),
\end{eqnarray}
where the MF spin propagator $D^{(0)}_{\varsigma}({\bf k},\omega)$ in the presence of the
electronic nematicity has been derived as \cite{Cao22},
\begin{eqnarray}
D^{(0)}_{\varsigma}({\bf k},\omega)&=&{B^{(\varsigma)}_{\bf k}
\over 2\omega^{(\varsigma)}_{\bf k}}
\left ( {1\over \omega-\omega^{(\varsigma)}_{\bf k}}
-{1\over\omega+\omega^{(\varsigma)}_{\bf k}}\right ), \label{MFSGF}
\end{eqnarray}
with the spin orthorhombic excitation spectrum $\omega^{(\varsigma)}_{\bf k}$, and the weight
function of the spin excitation spectrum $B^{(\varsigma)}_{{\bf{k}}}$ that have been given
explicitly in Ref. \onlinecite{Cao22}.

For the derivation of the electron diagonal and off-diagonal propagators, a full charge-spin
recombination scheme has been proposed based on the kinetic-energy-driven superconductivity
\cite{Feng15a}, where the coupling form between the electrons and a strongly dispersive spin
excitation is the same as that between the charge carriers and a strongly dispersive spin
excitation, i.e., the form of the self-consistent equations fulfilled by the electron diagonal
and off-diagonal propagators is the same as the form in Eq. (\ref{CCSCES}) fulfilled by the
charge-carrier diagonal and off-diagonal propagators. In this case, a charge carrier and a
localized spin in the fermion-spin representation (\ref{CSS}) are fully recombined into a
constrained electron in which the charge-carrier diagonal and off-diagonal propagators
$g_{\varsigma}({\bf k},\omega)$ and $\Gamma^{\dagger}_{\varsigma}({\bf k}, \omega)$ in
Eq. (\ref{CCSCES}) are replaced by the electron diagonal and off-diagonal propagators
$G_{\varsigma}({\bf k},\omega)$ and $\Im^{\dagger}_{\varsigma}({\bf k},\omega)$, respectively,
and then the electron diagonal and off-diagonal propagators of the $t$-$J$ model (\ref{tJ-model})
at zero magnetic field satisfy the following self-consistent equations,
\begin{subequations}\label{ESCES}
\begin{eqnarray}
G_{\varsigma}({\bf k},\omega)&=&G^{(0)}_{\varsigma}({\bf k},\omega)
+G^{(0)}_{\varsigma}({\bf k},\omega)[\Sigma^{(\varsigma)}_{\rm ph}({\bf k},\omega)
G_{\varsigma}({\bf k},\omega)\nonumber\\
&-&\Sigma^{(\varsigma)}_{\rm pp}({\bf k},\omega)\Im^{\dagger}_{\varsigma}({\bf k},\omega)],~~~~~~~
\label{EDGF} \\
\Im^{\dagger}_{\varsigma}({\bf k},\omega)&=&G^{(0)}_{\varsigma}({\bf k},-\omega)
[\Sigma^{(\varsigma)}_{\rm ph}({\bf k},-\omega)\Im^{\dagger}_{\varsigma}({\bf k},\omega)\nonumber\\
&+&\Sigma^{(\varsigma)}_{\rm pp}({\bf k},\omega)G_{\varsigma}({\bf k},\omega)], ~~~~~~
\label{EODGF}
\end{eqnarray}
\end{subequations}
where $G^{(0)}_{\varsigma}({\bf k},\omega)$ is the electron diagonal propagator of the $t$-$J$
model (\ref{tJ-model}) at zero magnetic field in the tight-binding approximation, and has been
obtained as \cite{Cao22},
\begin{eqnarray}\label{MFEGF}
G^{(0)}_{\varsigma}({\bf k},\omega)&=&{1\over \omega-\varepsilon^{(\varsigma)}_{\bf k}},
\end{eqnarray}
while the electron self-energies $\Sigma^{(\varsigma)}_{\rm ph}({\bf k},\omega)$ in the
particle-hole channel and $\Sigma^{(\varsigma)}_{\rm pp}({\bf k},\omega)$ in the
particle-particle channel can be obtained directly from the corresponding parts of the
charge-carrier self-energies $\Sigma^{(\rm h)}_{{\rm ph}}(\varsigma,{\bf k},\omega)$ in the
particle-hole channel and $\Sigma^{(\rm h)}_{{\rm pp}}(\varsigma,{\bf k},\omega)$ in the
particle-particle channel in Eq. (\ref{CCSE}) by the replacement of the full charge-carrier
diagonal and off-diagonal propagators
$g_{\varsigma}({\bf k},\omega)$ and $\Gamma^{\dagger}_{\varsigma}({\bf k},\omega)$
with the corresponding full electron diagonal and off-diagonal propagators
$G_{\varsigma}({\bf k},\omega)$ and $\Im^{\dagger}_{\varsigma}({\bf k},\omega)$ as,
\begin{subequations}\label{ESE}
\begin{eqnarray}
\Sigma^{(\varsigma)}_{\rm ph}({\bf k},i\omega_{n})&=&{1\over N^{2}}\sum_{{\bf p},{\bf p}'}
[\bar{\Lambda}^{(\varsigma)}_{{\bf p}+{\bf p}'+{\bf k}}]^{2}\nonumber\\
&\times&{1\over\beta}\sum_{ip_{m}}G_{\varsigma}({{\bf p}+{\bf k}},ip_{m}+i\omega_{n})
\Pi_{\varsigma}({\bf p},{\bf p}',ip_{m}),\label{ESE-ph}\nonumber\\
~~\\
\Sigma^{(\varsigma)}_{\rm pp}({\bf k},i\omega_{n})&=&{1\over N^{2}}\sum_{{\bf p},{\bf p}'}
[\bar{\Lambda}^{(\varsigma)}_{{\bf p}+{\bf p}'+{\bf k}}]^{2}\nonumber\\
&\times&{1\over \beta}\sum_{ip_{m}}\Im^{\dagger}_{\varsigma}({\bf p}+{\bf k},ip_{m}+i\omega_{n})
\Pi_{\varsigma}({\bf p},{\bf p}',ip_{m}),\nonumber\\
\end{eqnarray}
\end{subequations}
with the vertex function $\bar{\Lambda}^{(\varsigma)}_{\bf k}=4t[V^{(x)}_{\rm cor}(1-\varsigma)
\gamma_{{\bf k}_{x}}+V^{(y)}_{\rm cor}(1+\varsigma)\gamma_{{\bf k}_{y}}]-2t'(V^{(x)}_{\rm cor}
+V^{(y)}_{\rm cor})\gamma_{\bf{k}}'$, where the vertex correct in terms of $V^{(x)}_{\rm cor}$
and $V^{(y)}_{\rm cor}$ for the electron self-energies in the particle-hole and particle-particle
channels has been introduced for a better description of the nematic-order state strength
dependence of the electromagnetic response, which is different from the previous discussions of
the electronic structure of cuprate superconductors with coexisting electronic nematicity
\cite{Cao22,Cao21}, where this vertex correct is ignored. As in the previous discussions
\cite{Cao22,Cao21}, the electron self-energy in the particle-hole channel
$\Sigma^{(\varsigma)}_{\rm ph}({\bf k},\omega)$ represents the electron quasiparticle coherence,
while the electron self-energy in the particle-particle channel
$\Sigma^{(\varsigma)}_{\rm pp}({\bf k},\omega)$ represents the momentum and energy dependence of
the SC gap,
$\Sigma^{(\varsigma)}_{\rm pp}({\bf k},\omega)=\bar{\Delta}^{(\varsigma)}({\bf k},\omega)$.

In order to self-consistently determine all the parameters, the next step is to separate the
electron self-energy $\Sigma^{(\varsigma)}_{\rm ph}({\bf k},\omega)$ in the particle-hole
channel into its symmetric and antisymmetric parts as:
$\Sigma^{(\varsigma)}_{\rm ph}({\bf k},\omega)=\Sigma^{(\varsigma)}_{\rm phe}({\bf k},\omega)
+\omega\Sigma^{(\varsigma)}_{\rm pho}({\bf k},\omega)$. Following the common practice, this
antisymmetric part $\Sigma^{(\varsigma)}_{\rm pho}({\bf k},\omega)$ is defined as the electron
quasiparticle coherent weight: $Z^{(\varsigma)-1}_{\rm F}({\bf k},\omega)=1-{\rm Re}
\Sigma^{(\varsigma)}_{\rm pho}({\bf k},\omega)$. In an interacting electron system, everything
happens near the electron Fermi surface (EFS). As a case in low-energy close to EFS, the SC gap
and electron quasiparticle coherent weight can be discussed in the static-limit approximation,
\begin{subequations}\label{EPGF-EQCW}
\begin{eqnarray}
\bar{\Delta}^{(\varsigma)}({\bf k})&=&\bar{\Delta}^{(\varsigma)}_{\hat{x}}\gamma_{{\bf k}_{x}}
-\bar{\Delta}^{(\varsigma)}_{\hat{y}}\gamma_{{\bf k}_{y}},\label{EPGF}\\
{1\over Z^{(\varsigma)}_{\rm F}}&=&1-{\rm Re}
\Sigma^{(\varsigma)}_{\rm pho}({\bf k},\omega=0)\mid_{{\bf k}=[\pi,0]}, ~~~\label{EQCW}
\end{eqnarray}
\end{subequations}
where the wave vector ${\bf k}$ in $Z^{(\varsigma)}_{\rm F}({\bf k})$ has been chosen as
${\bf k}=[\pi,0]$ just as it has been done in the ARPES experiments \cite{DLFeng00,Ding01}.

With the help of the above static-limit approximation for $\bar{\Delta}^{(\varsigma)}({\bf k})$
and $Z^{(\varsigma)}_{\rm F}$ in Eq. (\ref{EPGF-EQCW}), the renormalized electron diagonal and
off-diagonal propagators are obtained from Eq. (\ref{ESCES}) as,
\begin{subequations}\label{MF-EGFS}
\begin{eqnarray}
G^{\rm (RMF)}_{\varsigma}({\bf k},\omega)&=&Z^{(\varsigma)}_{\rm F}
\left ({U^{(\varsigma)2}_{\bf k}\over \omega-E^{(\varsigma)}_{\bf k}}+{V^{(\varsigma)2}_{\bf k}
\over\omega +E^{(\varsigma)}_{\bf k}} \right ), \\
\Im^{{\rm (RMF)}\dagger}_{\varsigma}({\bf k},\omega)&=&-Z^{(\varsigma)}_{\rm F}
{\bar{\Delta}^{(\varsigma)}_{\rm Z}({\bf k})\over 2E^{(\varsigma)}_{\bf k}}\left ({1\over\omega
-E^{(\varsigma)}_{\bf k}}-{1\over\omega+E^{(\varsigma)}_{\bf k}}\right ),\nonumber\\
\end{eqnarray}
\end{subequations}
where the SC quasiparticle coherence factors $U^{(\varsigma)}_{\bf k}$ and
$V^{(\varsigma)}_{\bf k}$ are given explicitly by,
\begin{subequations}\label{EBCSCF}
\begin{eqnarray}
U^{(\varsigma)2}_{\bf k}&=&{1\over 2}\left ( 1+{\bar{\varepsilon}^{(\varsigma)}_{\bf k}\over
E^{(\varsigma)}_{\bf k}}\right ),\\
V^{(\varsigma)2}_{\bf k}&=&{1\over 2}\left ( 1-{\bar{\varepsilon}^{(\varsigma)}_{\bf k}\over
E^{(\varsigma)}_{\bf k}}\right ),
\end{eqnarray}
\end{subequations}
and fulfills the constraint $U^{(\varsigma)2}_{\bf k}+V^{(\varsigma)2}_{\bf k}=1$. In particular,
the renormalized electron diagonal and off-diagonal propagators in Eq. (\ref{MF-EGFS}) can be
also expressed explicitly in the Nambu representation as quoted in Eq. (\ref{NPEGF}).

Substituting these renormalized electron diagonal and off-diagonal propagators in
Eq. (\ref{MF-EGFS}) and MF spin propagator in Eq. (\ref{MFSGF}) into Eqs. (\ref{ESE}) and
performing the summation over bosonic Matsubara frequencies yield the final forms of the
electron self-energies in the particle-hole and particle-particle channels as,
\begin{widetext}
\begin{subequations}\label{ESE-1}
\begin{eqnarray}
{\Sigma}^{(\varsigma)}_{\rm ph}({\bf{k}},{\omega})&=&\frac{1}{N^{2}}
\sum_{{\bf{p}}{\bf{p}'}{\nu}}(-1)^{\nu+1}\bar{\Omega}^{(\varsigma)}_{{\bf{p}}{\bf{p}'}{\bf{k}}}
\left [ U^{(\varsigma)2}_{\bf{p}+\bf{k}}\left (
\frac{F^{(\varsigma)}_{1\nu}({\bf p},{\bf p}',{\bf k})}{\omega
+\omega^{(\nu)}_{\varsigma{\bf{p}}{\bf{p}}'}-E^{(\varsigma)}_{\bf{p}+\bf{k}}}
-\frac{F^{(\varsigma)}_{2\nu}({\bf p},{\bf p}',{\bf k})}{\omega-
\omega^{(\nu)}_{\varsigma{\bf{p}}{\bf{p}}'}-E^{(\varsigma)}_{\bf{p}+\bf{k}}} \right)\right.
\nonumber\\
&+& \left . V^{(\varsigma)2}_{\bf{p}+\bf{k}} \left (
\frac{F^{(\varsigma)}_{1\nu}({\bf p},{\bf p}',{\bf k})}{\omega-
\omega^{(\nu)}_{\varsigma{\bf{p}}{\bf{p}}'}+E^{(\varsigma)}_{\bf{p}+\bf{k}}}
-\frac{F^{(\varsigma)}_{2\nu}({\bf p},{\bf p}',{\bf k})}{\omega
+\omega^{(\nu)}_{\varsigma{\bf{p}}{\bf{p}}'}
+E^{(\varsigma)}_{\bf{p}+\bf{k}}}  \right )  \right ],\label{ph-ESE}\\
{\Sigma}^{(\varsigma)}_{\rm pp}({\bf{k}},{\omega})&=&\frac{1}{N^{2}}
\sum_{{\bf{p}}{\bf{p}'}{\nu}}(-1)^{\nu}
\bar{\Omega}^{(\varsigma)}_{{\bf{p}}{\bf{p}'}{\bf{k}}}
\frac{\bar{\Delta}_{\varsigma{\rm{Z}}}({\bf{p}}+{\bf{k}})}
{2E^{(\varsigma)}_{{\bf{p}}+\bf{k}}}\left [ \left (
\frac{F^{(\varsigma)}_{1\nu}({\bf p},{\bf p}',{\bf k})} {\omega
+\omega^{(\nu)}_{\varsigma{\bf{p}}{\bf{p}}'}-E^{(\varsigma)}_{\bf{p}+\bf{k}}}
-\frac{F^{(\varsigma)}_{2\nu}({\bf p},{\bf p}',{\bf k})}{\omega
-\omega^{(\nu)}_{\varsigma{\bf{p}}{\bf{p}}'}
-E^{(\varsigma)}_{\bf{p}+\bf{k}}} \right)\right. \nonumber\\
&-&\left. \left ( \frac{F^{(\varsigma)}_{1\nu}({\bf p},{\bf p}',{\bf k})} {\omega-\omega^{(\nu)}_{\varsigma{\bf{p}}{\bf{p}}'}+E^{(\varsigma)}_{\bf{p}+\bf{k}}}
-\frac{F^{(\varsigma)}_{2\nu}({\bf p},{\bf p}',{\bf{k}})}{\omega
+\omega^{(\nu)}_{\varsigma{\bf{p}}{\bf{p}}'}
+E^{(\varsigma)}_{\bf{p}+\bf{k}}}  \right )  \right ], \label{pp-ESE}
\end{eqnarray}
\end{subequations}
respectively, where $\nu=1,2$, $\bar{\Omega}^{(\varsigma)}_{{\bf{p}}{\bf{p}'}{\bf{k}}}
=Z^{(\varsigma)}_{\rm{F}}[\bar{\Lambda}^{(\varsigma)}_{{\bf{p}}+{\bf{p}'}+{\bf{k}}}]^{2}
B^{(\varsigma)}_{\bf{p}'}B^{(\varsigma)}_{\bf{p}+\bf{p}'}/(4\omega^{(\varsigma)}_{\bf{p}'}
\omega^{(\varsigma)}_{\bf{p}+\bf{p}'})$, and the weight functions,
\begin{subequations}
\begin{eqnarray}
F^{(\varsigma)}_{1\nu}({\bf{p}},{\bf{p}}',{\bf{k}})&=& n_{\rm{F}}
(E^{(\varsigma)}_{\bf{p}+\bf{k}})\{1+n_{B}(\omega^{(\varsigma)}_{\bf{p}'+\bf{p}})
+n_{B}[(-1)^{\nu+1}\omega^{(\varsigma)}_{\bf{p}'}]\}
+n_{B}(\omega^{(\varsigma)}_{\bf{p}'+\bf{p}})
n_{B}[(-1)^{\nu+1}\omega^{(\varsigma)}_{\bf{p}'}],\\
F^{(\varsigma)}_{2\nu}({\bf{p}},{\bf{p}}',{\bf{k}})&=& [1
-n_{\rm{F}}(E^{(\varsigma)}_{\bf{p}+\bf{k}})]\{1+n_{B}(\omega^{(\varsigma)}_{\bf{p}'+\bf{p}})
+n_{B}[(-1)^{\nu+1}\omega^{(\varsigma)}_{\bf{p}'}]\}
+n_{B}(\omega^{(\varsigma)}_{\bf{p}'+\bf{p}})
n_{B}[(-1)^{\nu+1}\omega^{(\varsigma)}_{\bf{p}'}],~~~~
\end{eqnarray}
\end{subequations}
with $n_{B}(\omega)$ and $n_{F}(\omega)$ that are the boson and fermion distribution functions,
respectively.
\end{widetext}

In the fermion-spin representation (\ref{CSS}), the SC gap parameter in real space can be
expressed as \cite{Feng15,Feng0306,Feng12,Feng15a},
\begin{eqnarray}\label{cooper-pair}
\Delta^{(\varsigma)}(l-l')&=&\langle C^{\dagger}_{l\uparrow}C^{\dagger}_{l'\downarrow}
-C^{\dagger}_{l\downarrow}C^{\dagger}_{l'\uparrow}\rangle\nonumber\\
&=&\langle h_{l\uparrow}h_{l'\downarrow}S^{+}_{l} S^{-}_{l'}-h_{l\downarrow}h_{l'\uparrow}
S^{-}_{l}S^{+}_{l'}\rangle.~~~~~
\end{eqnarray}
In the doped regime without an antiferromagnetic long-range order, the charge carriers move
in the background of the spin liquid state, where the spin correlation functions
$\langle S^{+}_{l}S^{-}_{l'}\rangle=\langle S^{-}_{l}S^{+}_{l'}\rangle=\chi_{l-l'}$. In this
case, the SC gap parameter in Eq. (\ref{cooper-pair}) can be expressed approximately as:
$\Delta^{(\varsigma)}(l-l')\approx -\chi_{l-l'}\Delta^{\rm (h)}_{\varsigma}(l'-l)$, with the
charge-carrier pair gap parameter $\Delta^{\rm (h)}_{\varsigma}(l'-l)=\langle h_{l'\downarrow}
h_{l\uparrow}-h_{l'\uparrow}h_{l\downarrow}\rangle$. On the other hand, the ARPES measurements
\cite{Damascelli03,Campuzano04,Fink07} have indicated that in the real space the SC gap and
pairing force have a range of one lattice spacing, which therefore shows that the components
of the SC gap parameter $\bar{\Delta}^{(\varsigma)}_{\hat{x}}$ and
$\bar{\Delta}^{(\varsigma)}_{\hat{y}}$ in Eq. (\ref{EPGF}) can be obtained approximately as,
\begin{eqnarray}\label{components}
\bar{\Delta}^{(\varsigma)}_{\hat{x}}\approx -\chi_{1\hat{x}}\Delta^{\rm (h)}_{\varsigma\hat{x}},
~~~~
\bar{\Delta}^{(\varsigma)}_{\hat{y}}\approx -\chi_{1\hat{y}}\Delta^{\rm (h)}_{\varsigma\hat{y}},
\end{eqnarray}
where the components of the charge-carrier pair gap parameter
$\Delta^{\rm (h)}_{\varsigma\hat{x}}$ and $\Delta^{\rm (h)}_{\varsigma\hat{y}}$, and the spin
correlation functions $\chi_{1\hat{x}}$ and $\chi_{1\hat{y}}$ have been obtained
self-consistently in Ref. \onlinecite{Cao22}. In this case, the electron quasiparticle coherent
weight $Z^{(\varsigma)}_{\rm F}$, the vertex correction parameters $V^{(x)}_{\rm cor}$ and
$V^{(y)}_{\rm cor}$, and the electron chemical potential $\mu$ satisfy following four
self-consistent equations,
\begin{widetext}
\begin{subequations}\label{SCE3}
\begin{eqnarray}
\frac{1}{Z^{(\varsigma)}_{\rm{F}}} &=& 1+\frac{1}{N^{2}}\sum_{{\bf{p}}{\bf{p}'}{\nu}}(-1)^{\nu+1}
\bar{\Omega}^{(\varsigma)}_{{\bf{p}}{\bf{p}'}{\bf k}_{\rm A}}\left (
\frac{F^{(\varsigma)}_{1\nu}({\bf p},{\bf p}',{\bf k}_{\rm A})}
{(\omega^{(\nu)}_{\varsigma{\bf{p}}{\bf{p}}'}-E^{(\varsigma)}_{\bf{p}+{\bf k}_{\rm A}})^{2}}
+\frac{F^{(\varsigma)}_{2\nu}({\bf{p}},{\bf{p}}',{\bf k}_{\rm A})}
{(\omega^{(\nu)}_{\varsigma{\bf{p}}{\bf{p}}'}+E^{(\varsigma)}_{\bf{p}+{\bf k}_{\rm A}})^{2}}
\right ), ~~~~~~~~~\\
\bar{\Delta}^{(\varsigma)}_{\hat{x}} &=& \frac{8}{N^{3}}
\sum_{{\bf{p}}{\bf{p}'}{\bf{k}}{\nu}}(-1)^{\nu}Z^{(\varsigma)}_{\rm{F}}
\bar{\Omega}^{(\varsigma)}_{{\bf{p}}{\bf{p}'}{\bf{k}}}
\frac{\gamma_{{\bf k}_{x}}(\bar{\Delta}^{(\varsigma)}_{\hat{x}}\gamma_{{\bf p}_{x}+{\bf k}_{x}}
-\bar{\Delta}^{(\varsigma)}_{\hat{y}}\gamma_{{\bf p}_{y}+{\bf k}_{y}})}
{E^{(\varsigma)}_{\bf{p}+\bf{k}}}\left (
\frac{F^{(\varsigma)}_{1\nu}({\bf{p}},{\bf{p}}',{\bf{k}})}
{\omega^{(\nu)}_{\varsigma{\bf{p}}{\bf{p}}'}
-E^{(\varsigma)}_{\bf{p}+\bf{k}}}-\frac{F^{(\varsigma)}_{2\nu}({\bf{p}},{\bf{p}}',{\bf{k}})}
{\omega^{(\nu)}_{\varsigma{\bf{p}}{\bf{p}}'}+E^{(\varsigma)}_{\bf{p}+\bf{k}}} \right ),\\
\bar{\Delta}^{(\varsigma)}_{\hat{y}} &=& \frac{8}{N^{3}}
\sum_{{\bf{p}}{\bf{p}'}{\bf{k}}{\nu}}(-1)^{\nu+1}Z^{(\varsigma)}_{\rm{F}}
\bar{\Omega}^{(\varsigma)}_{{\bf{p}}{\bf{p}'}{\bf{k}}}
\frac{\gamma_{{\bf k}_{y}}(\bar{\Delta}^{(\varsigma)}_{\hat{x}}\gamma_{{\bf p}_{x}+{\bf k}_{x}}
-\bar{\Delta}^{(\varsigma)}_{\hat{y}}\gamma_{{\bf p}_{y}+{\bf k}_{y}})}
{E^{(\varsigma)}_{\bf{p}+\bf{k}}}\left (
\frac{F^{(\varsigma)}_{1\nu}({\bf{p}},{\bf{p}}',{\bf{k}})}
{\omega^{(\nu)}_{\varsigma{\bf{p}}{\bf{p}}'}
-E^{(\varsigma)}_{\bf{p}+\bf{k}}}-\frac{F^{(\varsigma)}_{2\nu}({\bf{p}},{\bf{p}}',{\bf{k}})}
{\omega^{(\nu)}_{\varsigma{\bf{p}}{\bf{p}}'}+E^{(\varsigma)}_{\bf{p}+\bf{k}}} \right ),\\
1-\delta &=&\frac{1}{2N}\sum_{{\bf{k}}}Z^{(\varsigma)}_{\rm{F}} \left(
1-\frac{\bar{\varepsilon}^{(\varsigma)}_{\bf{k}}}{E^{(\varsigma)}_{{\bf{k}}}}{\rm{tanh}}
\left[ \frac{1}{2}\beta{E^{(\varsigma)}_{\bf{k}}} \right ] \right ),\label{SCE-EFS}
\end{eqnarray}
\end{subequations}
\end{widetext}
where ${\bf k}_{\rm A}=[\pi,0]$. The above self-consistent equations (\ref{SCE3}) have been
solved numerically on a $120\times 120$ lattice in momentum space as our previous discussions
\cite{Cao22,Cao21}, and then the electron quasiparticle coherent weight $Z^{(\varsigma)}_{\rm F}$,
the vertex correction parameters $V^{(x)}_{\rm cor}$ and $V^{(y)}_{\rm cor}$, and the electron
chemical potential $\mu$ are obtained self-consistently. In particular, at the condition of the
SC gap parameter $\bar{\Delta}^{(\varsigma)}=0$ [then $\bar{\Delta}^{(\varsigma)}_{\hat{x}}=0$
and $\bar{\Delta}^{(\varsigma)}_{\hat{y}}=0$], the evolution of $T_{\rm c}$ with the
nematic-order state strength at a given doping can be also determined self-consistently from
the above self-consistent equations (\ref{SCE3}).

\end{appendix}

\end{document}